\documentclass[apj,twocolumn,breaklinks,colorlinks,citecolor=blue,urlcolor=blue]{openjournal}

\usepackage{amsmath}
\usepackage{xcolor}
\colorlet{RED}{red}

\usepackage{orcidlink}
\newcommand{\newNZR}[1]{#1}

\newcommand{\edit}[1]{#1}

\begin{document}

\title{Finding the unusual red giant remnants of cataclysmic variable mergers}

\author{\vspace{-1.4cm}Nicholas Z. Rui\,\orcidlink{0000-0002-1884-3992}$^{1}$}
\author{Jim Fuller\,\orcidlink{0000-0002-4544-0750}$^{1}$}

\affiliation{$^1$TAPIR, California Institute of Technology, Pasadena, CA 91125, USA}


 
\begin{abstract}
Mergers between helium white dwarfs and main-sequence stars are likely common, producing red giant-like remnants making up roughly a few percent of all low-mass ($\lesssim2M_\odot$) red giants.
Through detailed modeling, we show that these merger remnants possess distinctive photometric, asteroseismic, and surface abundance signatures through which they may be identified.
During hydrogen shell burning, merger remnants reach higher luminosities and possess pulsations which depart from the usual degenerate sequence on the asteroseismic $\Delta\nu$--$\Delta\Pi$ diagram for red giant branch stars.
For sufficiently massive helium white dwarfs, merger remnants undergo especially violent helium flashes which can dredge up a large amount of core material (up to $\sim0.1M_\odot$) into the envelope.
Such post-dredge-up remnants are more luminous than normal red clump stars, are surface carbon-, helium-, and possibly lithium-rich, and possess a wider range of asteroseismic g-mode period spacings and mixed-mode couplings.
Recent asteroseismically determined low-mass ($\lesssim0.8M_\odot$) red clump stars may be core helium-burning remnants of mergers involving lower-mass helium white dwarfs.
\end{abstract}



\section{Introduction} \label{sec:intro}

Growing evidence suggests that mergers between main-sequence (MS) stars and low-mass white dwarfs should be fairly common \edit{\citep{schreiber2015three}}.
The likely outcome of such a merger is to accrete the MS star onto the white dwarf, igniting hydrogen shell burning and creating an unusual red giant (RG) star.
These RG remnants of those mergers should exist within the stellar population, and may be identifiable through a combination of their photometry, pulsations, and surface abundances.

Cataclysmic variables, or CVs, are stably mass-transferring systems with white dwarf (WD) accretors and MS donors. 
Typically, the progenitor binaries of CVs are post-common-envelope systems \edit{\citep{paczynski1976structure,belczynski2005new,toonen2013effect,camacho2014monte,ablimit2016monte,zorotovic2022close}} which have been tightened by magnetic braking and gravitational waves until the onset of Roche lobe overflow \citep{knigge2011evolution}.

For a while, the research field has been dogged by a \textit{WD mass problem} \citep[e.g.,][]{zorotovic2020cataclysmic}:
\begin{enumerate}
    \item the typical accretor mass in CVs is $\simeq~0.8M_\odot$, greater than the typical \edit{observed} mass of carbon--oxygen white dwarfs (CO WDs) $\simeq~0.6M_\odot$, and
    \item \edit{only a small handful of} helium(-core) white dwarf (hereafter He WD) accretors have ever been discovered in \edit{accreting systems}, even though they should be frequent\edit{, observable} outcomes of conventional binary stellar evolution \citep{zorotovic2011post,zorotovic2020cataclysmic,pala2022constraining}.
\end{enumerate}

Where are all of the low-mass CV accretors?

\begin{figure*}
    \centering
    \includegraphics[width=\textwidth]{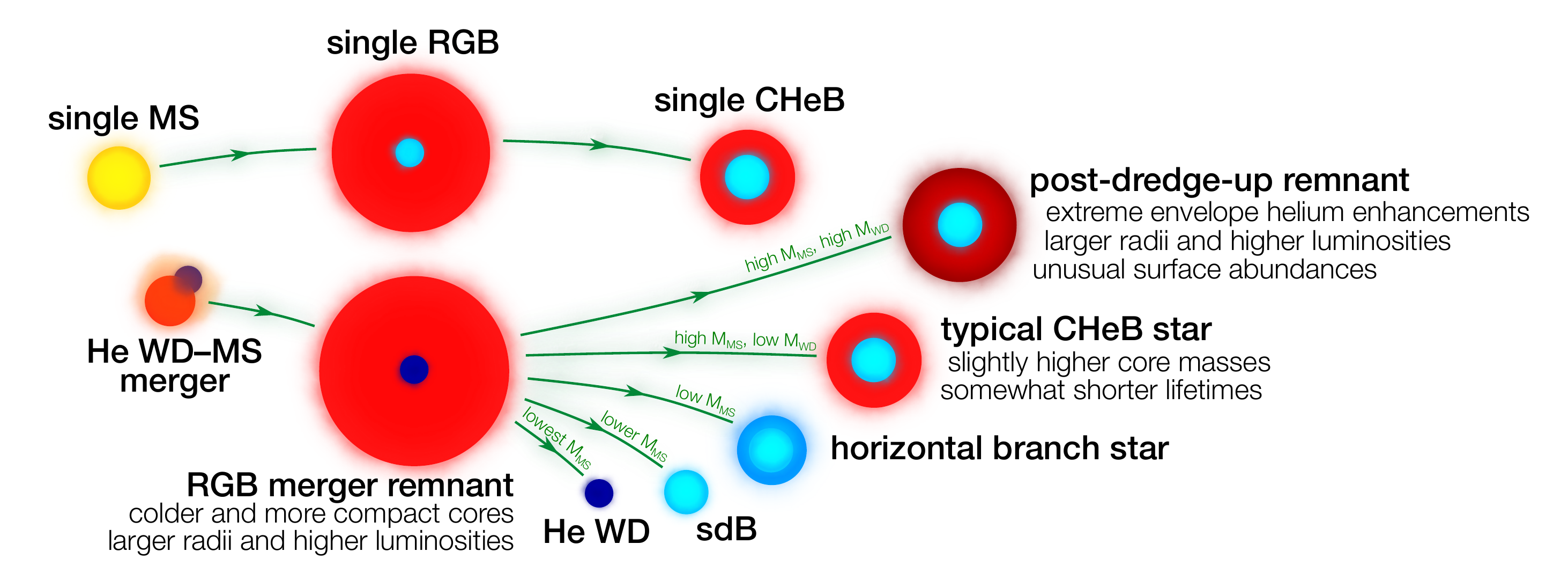}
    \caption{A summary diagram of the behavior of the He WD--MS merger remnants investigated in this work.
    On the RGB, merger remnants have cold and slightly more compact cores, which inflates their radii and changes the radiative structures of their cores to which asteroseismology is sensitive (Section \ref{rgb}).
    \edit{Merger remnants which are sufficiently massive enough to reach the tRGB} ignite helium in abnormally energetic helium flashes.
    For massive enough progenitor He WDs, these events can dredge up significant amounts of helium into merger remnants' envelopes (Section \ref{chebsupersect}), changing their surface abundances, and greatly modifying both their structure and evolution.
    \edit{In our models, such core dredge-up events occur when $M_{\mathrm{WD}}\geq0.27M_\odot$, although the exact threshold depends on other factors such as the cooling age of the He WD (see Appendix \ref{coolingage}).}
    Less massive remnants become horizontal branch stars (Section \ref{horizontalbranch}), hot subdwarfs, or He WDs (not explored in this work).}
    \label{fig:hewd_ms_cartoon}
\end{figure*}

In the past decade, it was realized that the WD mass problem can naturally be solved by an extra, accretor mass-dependent angular momentum loss mechanism \citep{schreiber2015three}.
This so-called \textit{consequential angular momentum loss} would preferentially destabilize the mass transfer of CV-like systems with lower-mass white dwarfs, causing them to quickly merge before they can be observed \citep{belloni2018no}.
Physically, this mechanism may be a frictional effect associated with nova events \edit{\citep{shen2015every,shen2022binary}}, which are expected to be much longer-duration for lower-mass white dwarfs \citep{shara1993multiple} \edit{and may cause mergers to occur on the timescale of hours \citep{shen2015every}}, although the subsequent analysis is agnostic to the details.

If CVs with low accretor masses are missing because they merge, it is obvious to ask whether their remnants can be observed and identified.
In the case where the progenitor system is a close He WD--MS binary (the scenario of focus in this study), the remnant is expected to evolve along highly modified versions of the red giant branch (RGB) and possibly core helium burning (CHeB, or the red clump) phases of isolated, low-mass ($M~\lesssim~2M_\odot$) stars.

More recently, in only the last few years, the community has realized that stellar interactions may produce RGs with conspicuous and lasting asteroseismic signatures.
\citet{rui2021asteroseismic} and \citet{deheuvels2022seismic} show that many first-ascent RGs which gain hydrogen mass from companions (either through merger or stable mass transfer) possess unusually low gravity-mode period spacings.
\citet{li2022discovery} and \citet{matteuzzi2023red} similarly identify anomalously undermassive CHeB stars through asteroseismology.
Because asteroseismology probes \textit{internal} structures, it constrains separate information from traditional techniques which probe the surface properties of the star.
It may thus play a critical role in identifying the remnants of cataclysmic variable mergers.

In this work, we investigate observable signatures of He WD--MS merger remnants.
The physical picture we advance is shown schematically in Figure \ref{fig:hewd_ms_cartoon}.
We construct and evolve merger remnant models (described in Section \ref{models}), and demonstrate a significant number of telltale signs (photometric, asteroseismic, and surface compositional) which are complex downstream consequences of a highly cold and degenerate core during hydrogen shell burning.
Sections \ref{rgb} and \ref{chebsupersect} discuss merger remnants during hydrogen shell burning (RGB) and helium core burning (CHeB), respectively.
\newNZR{In Section \ref{candidates}, we identify some merger remnant candidates in existing observations.
Section \ref{discussion} discusses other possible signatures, and contextualizes this work in the broader nascent field of binary interaction asteroseismology.}
Section \ref{conclusion} summarizes our key findings.

\section{Stellar models} \label{models}

By the time of merger, the He WD component of a close He WD--MS binary has undergone an extended phase of radiative cooling lasting potentially up to gigayears.
When the binary subsequently merges, the MS component (now the remnant's envelope) quickly ignites hydrogen in a burning shell around the He WD (now the remnant's core) and sets up thermal equilibrium on a short envelope thermal timescale:
\begin{equation} \label{thenv}
    \begin{split}
        \tau_{\mathrm{th,env}} &= \frac{GM^2}{RL} \\
        &\!\!\!\!\!\!\!\!\!\approx 0.1\,\mathrm{Myr}\times\left(\frac{M}{M_\odot}\right)^2\left(\frac{R}{10R_\odot}\right)^{-1}\left(\frac{L}{30L_\odot}\right)^{-1}
    \end{split}
\end{equation}

In contrast, the degenerate core only thermalizes (via electron-mediated conduction) with the hydrogen burning shell
on a much longer timescale $\sim\!10$--$100\,\mathrm{Myr}$ comparable to the duration of the RGB phase (see Appendix \ref{coreth}).
Therefore, as previously noticed by \citet{zhang2017evolution}, the low-entropy core can persist for long enough to influence the long-term evolution of the remnant.

Using Modules for Experiments in Stellar Astrophysics \citep[MESA, version r10398;][]{paxton2010modules,paxton2013modules,paxton2015modules,paxton2018modules}, we create and evolve evolutionary models of merger remnants, consisting of cold helium cores surrounded by hydrogen envelopes\footnote{Inlists and other files required to reproduce our results can be found at the following Zenodo link: \href{https://zenodo.org/records/10828187}{https://zenodo.org/records/10828187}}.

\begin{figure*}
    \centering
    \includegraphics[width=\textwidth]{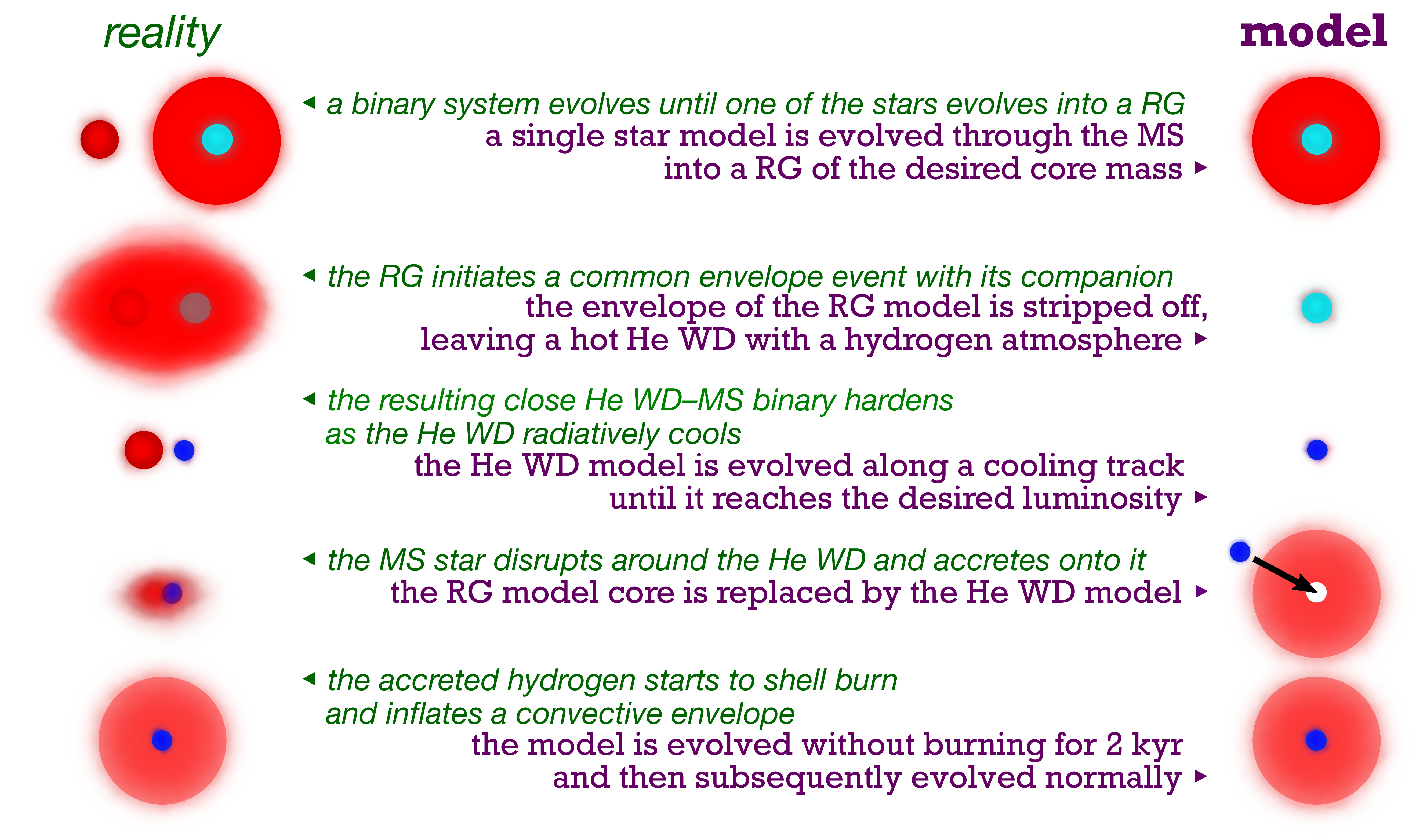}
    \caption{\edit{A summary of the evolution of one possible progenitor system of a He WD--MS merger remnant (\textit{left}), juxtaposed against our procedure for constructing the stellar model of a merger remnant in MESA (\textit{right}).}}
    \label{fig:model_construction}
\end{figure*}

We do this in the following stages \edit{(summarized in Figure \ref{fig:model_construction})}:
\begin{enumerate}
    \item An isolated star with $M=1.2M_\odot$ is evolved through the MS and part of the RGB until it attains the desired initial core mass of the remnant, $M_{\mathrm{WD}}$.
    \newNZR{This initial mass is chosen to be close to typical RG masses from \textit{Kepler} \citep[e.g.,][]{yu2018asteroseismology}.
    Because \edit{single} RGs with $M\lesssim2M_\odot$ all obey the same core mass--luminosity relation, the subsequent analysis is insensitive to this choice.}
    
    \item The hydrogen envelope is removed by applying a high mass-loss rate (using \texttt{relax\_initial\_mass} with $|\dot{M}|=10^{-3}M_\odot\,\mathrm{yr}^{-1}$).
    To simulate a realistic white dwarf atmosphere, we retain an additional $M_{\mathrm{atm}}=10^{-4}M_\odot$ of material from the base of the envelope, on top of the core mass $M_{\mathrm{WD}}$. 
    
    \item The resulting object (which quickly relaxes into a He WD) is evolved through a cooling track
    to $\log(L_{\mathrm{WD}}/L_\odot)=-4.0$.

    \item To stellar-engineer the merger remnant, we start with a scaffold RG model with core mass $M_{\mathrm{WD}}$.
    To build the scaffold model, we start with the original RG model from above\edit{, and} remove envelope mass from the scaffold model via relaxation with $|\dot{M}|=10^{-3}M_\odot\,\mathrm{yr}^{-1}$ with \edit{nuclear} burning disabled, until the RG attains the desired final envelope mass $M_{\mathrm{MS}}$ (physically equal to the mass of the MS star participating in the merger, assuming no mass lost).

    \edit{Next, we} replace shells with mass coordinates $m\leq M_{\mathrm{WD}}$ with the He WD model\edit{, excluding the He WD's atmosphere (shells where $X\geq10^{-4}$).}
    \edit{Because the modified RG model's core is now more compact, we recalculate the radial coordinate grid to be consistent with $\mathrm{d}m=4\pi r^2\rho\,\mathrm{d}r$.}
    \newNZR{This produces a merger remnant with total mass $M_{\mathrm{tot}}=M_{\mathrm{MS}}+M_{\mathrm{WD}}$.}
    
    \newNZR{For most merger remnant models in this work, we \edit{somewhat} arbitrarily fix $M_{\mathrm{MS}}$ such that $M=0.8M_\odot$, to resemble fairly typical low-mass RGs which may result from CV mergers.}
    
    \item To ensure numerical convergence, the resulting remnant is evolved for $2\,\mathrm{kyr}$ without burning and gold tolerances disabled.
    Nuclear reactions and gold tolerances are re-enabled at $2\,\mathrm{kyr}$ and $4\,\mathrm{kyr}$, respectively.
    \edit{Although not initially so, the merger remnant model quickly reaches hydrostatic equilibrium.}
    Evolution through the helium flash involves disabling gold tolerances again, 
    and the models are terminated after helium-burning when the central helium fraction drops to $Y_c\leq10^{-3}$.
\end{enumerate}

In post-processing, we truncate the first $1\,\mathrm{Myr}$ (a few envelope thermal times) of our merger remnant models in order to avoid possibly unphysical transient behavior closely following the merger, which is not modelled accurately.

The relatively low He WD luminosity $L_{\mathrm{WD}}=10^{-4.0}L_\odot$ is chosen to highlight the effects of a highly degenerate core in the limiting case.
In Appendix \ref{coolingage}, we show that, once the core has thermalized sufficiently long on the RGB, the effect of increasing $L_{\mathrm{WD}}$ is very similar to that of decreasing $M_{\mathrm{WD}}$.
This is because these changes ultimately affect the entropy of the merger remnants' cores in the same direction, and subtler differences in the remnants' core temperature profiles are erased by thermal conduction on the core's thermal time (see Appendix \ref{coreth}).

Our models include the predictive mixing scheme described in Section 2.1 of \citet{paxton2018modules} to account for near-core mixing, and to suppress numerical instability associated with definition of the convective core boundary, especially during the CHeB phase.
We omit winds in order to avoid sensitivity to the wind prescription, which is highly uncertain.
The effect of winds at the tip of the RGB (or mass loss during the merger itself) is primarily to increase the value of $M_{\mathrm{MS}}$ required for a given merger remnant mass during CHeB.
Stellar models are initialized to solar metallicity, using the metal mass fractions given by \citet{grevesse1998standard}.
We use the built-in \texttt{pp\_cno\_extras\_o18\_ne22} network for nuclear reactions.

When evolving the non-merged and merger remnant models, we run MESA in the hydrodynamical mode (evolving the radial velocity variable $v_r$ explicitly), in order to stably evolve our models through the helium flash.
We find that this is particularly necessary in the merger remnant models, where the helium flash is abnormally violent (see Section \ref{chebsupersect}).
We are able to model merger remnants 
involving He WD masses as high as $M_{\mathrm{WD}}=0.38M_\odot$.
Higher He WD masses cause numerical problems during the He flash.

This scheme for producing He WD--MS merger remnants produces very similar models to those of \citet{zhang2017evolution}, who instead manually add the hydrogen envelope back onto the He WD using a large, time-dependent mass gain rate.
We find that our method more reliably produces a model which MESA can evolve without case-by-case human intervention, and allows for the successful evolution of remnants with more degenerate cores.

In order to demonstrate the effects of the enhanced core degeneracy in our merger remnant models, we also run a model (hereafter the \textit{non-merged} model) of a RG whose envelope has only partially been removed on the RGB (so that it has a total mass $M\!=\!0.80M_\odot$).
Besides having possibly a low-mass envelope, this model otherwise behaves like a normal RG and should be thought of as representing standard stellar evolution.

\section{Red giant branch} \label{rgb}

\begin{figure*}
    \centering
    \includegraphics[width=\textwidth]{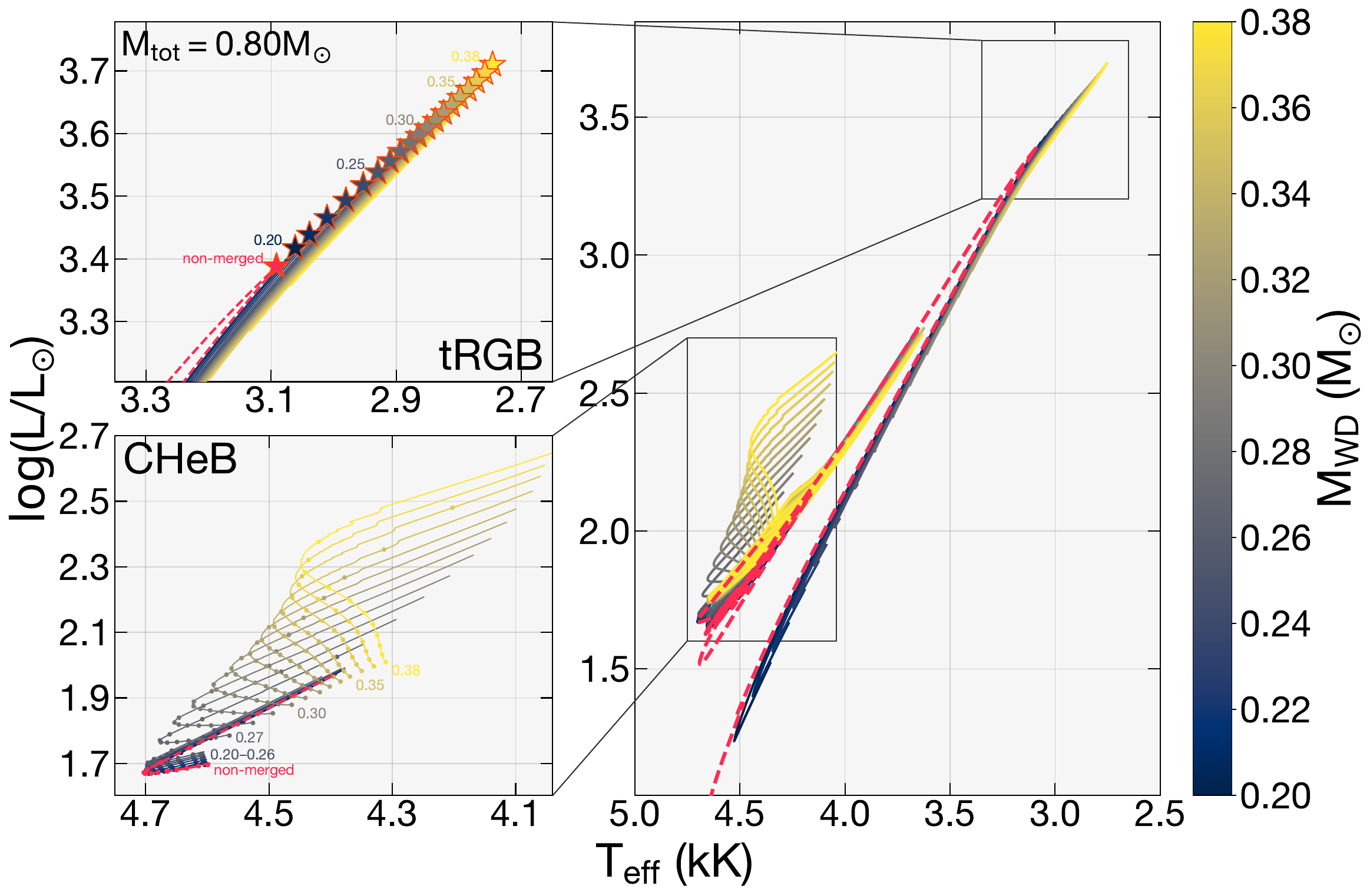}
    \caption{The trajectories on the Hertzsprung--Russell diagram of merger remnant models with varying He WD mass $M_{\mathrm{WD}}$ and total mass \edit{$M_{\mathrm{tot}}=M_{\mathrm{MS}}+M_{\mathrm{WD}}=0.80M_\odot$}.
    The \textit{red dashed} line indicates a non-merged model with $M=0.80M_\odot$.
    The \textit{top-left} and \textit{bottom-left} panels zoom into the tRGB and CHeB stages, respectively.
    The \textit{red-outlined} star symbols on the top-left panel indicate the location of the helium flash.
    On the \textit{bottom-left} panel, only evolution $\gtrsim5\,\mathrm{Myr}$ following the helium flash is shown, and \textit{points} are distributed $10\,\mathrm{Myr}$ apart.
    We have excised times near the helium flash\newNZR{, when the envelope sometimes becomes hydrodynamical.}
    }
    \label{fig:hrd_flashclump}
\end{figure*}

Soon after merger, a merger remnant quickly relaxes into a RG which behaves similarly to a normal star on the RGB.
Specifically, it is composed of an inert degenerate helium core surrounded by a tenuous hydrogen envelope which is inflated to a large radius by hydrogen burning in a thin shell at its base.
However, merger remnants differ from single RGs because their cores are cold (owing to the potentially long cooling phase of the progenitor He WD), rather than being almost isothermal with the burning shell.

Soon after their formation, merger remnants, especially those involving lower-mass He WDs, \newNZR{are out of thermal equilibrium and} temporarily shrink in radius. 
The duration of this phase is a decreasing function with $M_{\mathrm{WD}}$, lasting $\simeq10\,\mathrm{Myr}$ for $M_{\mathrm{WD}}=0.20M_\odot$ and dropping to $\lesssim1\,\mathrm{Myr}$ by $M_{\mathrm{WD}}=0.27M_\odot$.
The degree of this secular dimming also drops off strongly with $M_{\mathrm{WD}}$.
However, after this short-lived period, the remnant evolves monotonically up a modified version of the RGB.

Overall, this cold core has three consequences on the RGB:
\begin{enumerate}
    \item At fixed helium core mass, the star is brighter and has a larger radius (Section \ref{over-inflatedrgb}).
    Because the helium core of a remnant is colder and thus slightly more compact, the overlying hydrogen-burning shell has a significantly higher luminosity.

    \item The degenerate core alters the structure of the gravity-mode (g-mode) cavity (Section \ref{rgbdeltapg}), modifying the propagation of g modes in an observable way.
    
    \item The helium flash, which occurs when some shell in the core reaches a sufficient temperature $T\simeq10^8\,\mathrm{K}$, is slightly delayed (Section \ref{trgb}).
    This allows merger remnants to exceed the maximum luminosity attainable by a single RG at the tip of the RGB.
    Additionally, the more violent helium flash can have significant consequences for the CHeB stage (Section \ref{chebsupersect}).
\end{enumerate}


\subsection{More compact cores inflate the radius} \label{over-inflatedrgb}

In a typical RG, the total luminosity $L$ is essentially entirely determined by its core mass $M_c$ \citep{refsdal1970core}, with very little sensitivity to the mass of the envelope.
This correspondence is called the \textit{core mass--luminosity relation}.
Photometrically, this means that single RGs with similar core masses will appear essentially indistinguishable on a Hertzsprung--Russell diagram, even if their envelopes differ significantly in mass.
Asteroseismically, it results in a tight relationship between the large frequency spacing $\Delta\nu$ and the g-mode period spacing $\Delta\Pi$ \newNZR{(defined and interpreted below)} for degenerate-core RGs with $M\lesssim2M_\odot$ \citep[e.g.,][]{deheuvels2022seismic}.

Crucial to this relation is that the radius of the burning shell (which is similar to the radius of the core, $R_c$) is a strict function of $M_c$ (for a degenerate core, roughly $R_c \propto M_c^{-1/3}$).
In merger remnants, the core is cooler than the hydrogen-burning shell and therefore slightly more compact \citep{althaus2005mass}.
In this case, the luminosity 
is no longer fixed by $M_c$ alone, although it is still  determined by the environment around the burning shell.

In addition to modifying its placement on the Hertzsprung--Russell diagram (Figure \ref{fig:hrd_flashclump}), the inflated radius significantly modifies the observable pressure (p) modes in the envelope.
Asteroseismically, RGs are solar-like oscillators \citep{chaplin2013asteroseismology}, for which a frequency of maximum power $\nu_{\mathrm{max}}$ and large (p-mode) frequency spacing $\Delta\nu$ can be measured.
These observables are approximately related to the mass $M$ and radius $R$ of the RG as
\begin{equation} \label{numax}
    \nu_{\mathrm{max}} \propto g/T_{\mathrm{eff}}^{1/2} \propto M\,R^{-2}\,T_{\mathrm{eff}}^{-1/2}
\end{equation}

\noindent and
\begin{equation} \label{dnu}
    \Delta\nu \simeq \sqrt{G\Bar{\rho}} \propto M^{1/2}\,R^{-3/2}
\end{equation}

\noindent \citep{ulrich1986determination,brown1991detection}.

In the absence of additional information, $M_c$ and $R_c$ are not known.
While $L$, $\nu_{\mathrm{max}}$, and $\Delta\nu$ are all significantly different in merger remnants, this is only a direct consequence of their inflated radii.
Therefore, these observables alone cannot distinguish merger remnants except near the tip of the RGB (Section \ref{trgb}).
However, independent probes of the core, most notably the asteroseismic g-mode period spacing (Section \ref{rgbdeltapg}) can help distinguish merger remnants from normal stars.

\subsection{Asteroseismic signatures on the red giant branch} \label{rgbdeltapg}

Asteroseismically, RGs which are sufficiently low on the RGB \citep[roughly before the red bump;][]{pinccon2020probing} have strong enough mixed-mode coupling such that the g-mode period spacing $\Delta\Pi$ can be measured:
\begin{equation} \label{dpgformula}
    \Delta\Pi \approx \frac{\sqrt{2}\pi^2}{\int_{\mathcal{R}}(N/r)\,\mathrm{d}r}\mathrm{,}
\end{equation}

\noindent defined here specifically for the dipole 
($\ell=1$) modes.
This quantity probes the radiative core of the star.
The integral in Equation \ref{dpgformula} is taken over $\mathcal{R}$, the g-mode cavity of the RG, i.e., the region where \edit{$2\pi\nu_{\mathrm{max}}<N$}, where $N$ is the Brunt--V\"ais\"al\"a frequency.
At present, $\Delta\Pi$ has been measured for a few thousand RGs \citep[e.g.,][]{vrard2016period}, observationally constraining the structures of their radiative cores.


By Equation \ref{dpgformula}, more stratified radiative zones with larger integrals over $N\,\mathrm{d}\ln r$ have lower period spacings.
In Appendix \ref{degenbrunt}, we show that the Brunt--V\"ais\"al\"a frequency in the degenerate part of the radiative core is given by
\begin{equation}
\label{eq:Napprox}
    N^2 \approx N_0^2\frac{k_BT}{ZE_F}\left(1 - \frac{5}{2}\nabla\right) \propto \frac{T}{ZE_F}\left(1 - \frac{5}{2}\nabla\right)\mathrm{.}
\end{equation}

\noindent where $E_F$ is the Fermi energy of the core, $\nabla = d \ln T/d\ln P$, and the normalization $N_0^2$ is comparable to the dynamical frequency of the core, and is given by
\begin{equation}
    N_0^2 = \frac{\rho g^2}{p} \sim \sqrt{GM_c/R_c^3}\mathrm{.}
\end{equation}

Both a typical RG core and cooling WD are approximately isothermal ($\nabla\ll1$).
Then, ignoring the temperature dependence of the stellar structure for the moment, we approximately expect
\begin{equation}
    N \propto \sqrt{T}\mathrm{,}
\end{equation}
A ``cold,'' highly degenerate isothermal core will have a significantly smaller $N$, and therefore significantly larger $\Delta\Pi$, than a ``warm'' one \citep{bildsten1995nonradial}.
However, if there is a temperature gradient, it may significantly affect $\Delta\Pi$.

\begin{figure*}
    \centering
    \includegraphics[width=\textwidth]{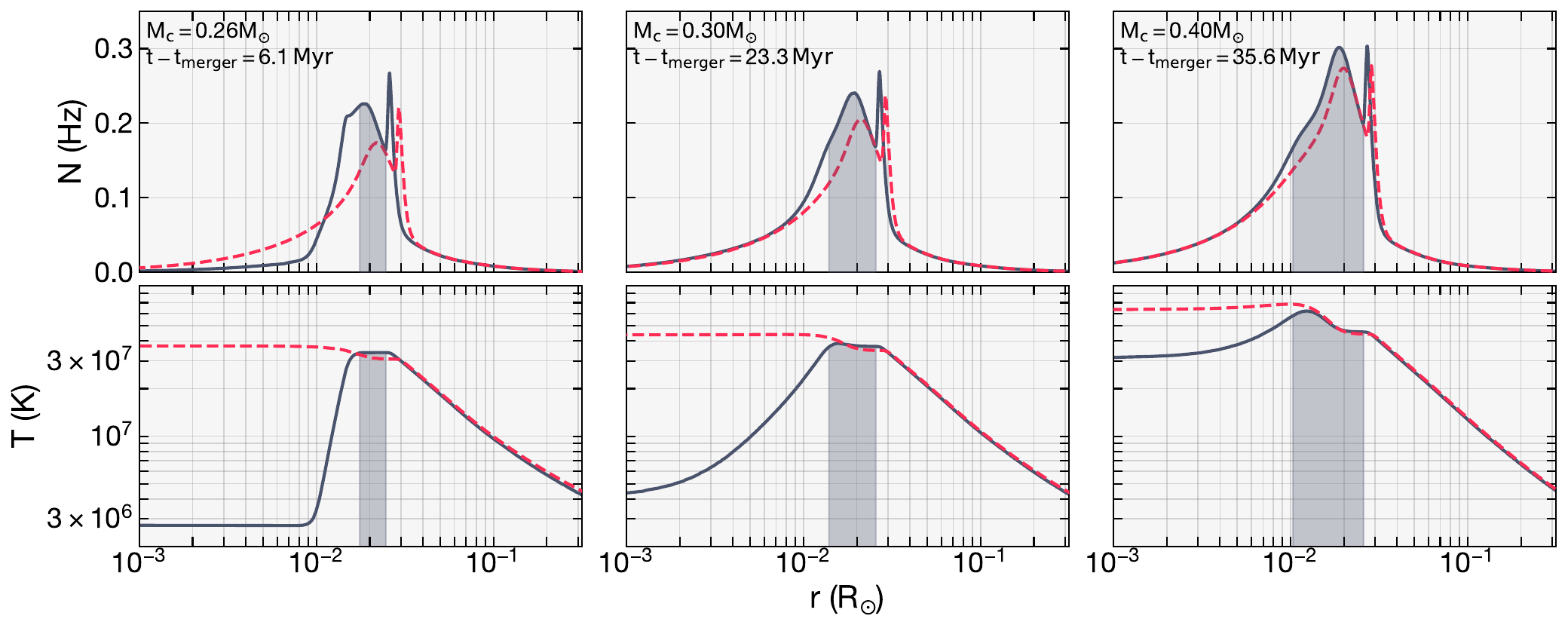}
    \caption{
    The Brunt--V\"ais\"al\"a frequency $N$ (\textit{top}) and temperature $T$ (\textit{bottom}) shown for the $M_{\mathrm{MS}}=0.55M_\odot$, $M_{\mathrm{WD}}=0.25M_\odot$ merger remnant (\textit{solid gray-blue}) and non-merged model of $M=0.80M_\odot$ (\textit{red dashed}), at various stages of evolution.
    The two models are compared at equal core masses $M_c$.
    The shaded region denotes helium \edit{deposited onto} the helium core of the merger remnant model \edit{by hydrogen burning} during its evolution.}
    \label{fig:explain_period_spacings}
\end{figure*}

The evolution of the buoyancy profile of a merger remnant with $M_{\mathrm{WD}}=0.25M_\odot$ and $M_{\mathrm{MS}}=0.55M_\odot$ (together with a non-merged model of equal mass) is shown in the top panels of Figure \ref{fig:explain_period_spacings}.
In He WD--MS merger remnants, the period spacing is modified by two effects which act roughly in opposite directions with comparable magnitudes.

First, at early times (left panels of Figure \ref{fig:explain_period_spacings}), the core of a merger remnant is roughly isothermal at its initial temperature $T\ll T_{\mathrm{shell}}$.
In comparison, the core of a single RG is roughly isothermal with the hydrogen-burning shell ($T\approx T_{\mathrm{shell}}$).
During this time period, the Brunt--V\"ais\"al\"a frequency profile in the core is suppressed from that of a single star by a factor $\sqrt{T_{\mathrm{shell}}/T}$.
This effect tends to increase the period spacing.

However, in a short time $<\tau_{\mathrm{th,core}}$, electronic thermal conduction (approximately obeying the nonlinear law in Equation \ref{nonlinear}) sets up a temperature gradient $\nabla$.
Curiously, we find that this temperature gradient happens to mostly cancel out the factor $\sqrt{T_{\mathrm{shell}}/T}$ (see Equation \ref{eq:Napprox}) to cause the $N$ profiles of the single and merger remnant models to be similar (\textit{top center} panels of Figure \ref{fig:explain_period_spacings}).
Equality between the Brunt--V\"ais\"al\"a frequency profiles occurs long before the interior of the core has thermalized to the shell temperature.
Therefore, in the relatively short time that it takes heat conduction to set up a temperature gradient throughout the core, the lower temperature of the merger remnant core no longer works to increase $\Delta\Pi$.
The reason for this ``coincidental'' cancellation
remains mysterious, but may be related to some properties of long-lived pre-thermalized solutions of nonlinear heat diffusion. These solutions \citep[called ``intermediate asymptotics'';][]{barenblatt1996scaling} exhibit self-similar behavior in many nonlinear heat diffusion problems (often in the context of diffusion in a porous medium), and have been studied extensively in applied mathematics \citep[by, e.g.,][]{witelski1998self,galaktionov2004intermediate,hayek2014exact}.


The second effect is that, because $N$ scales with the dynamical frequency of the core, it is higher in a more compact core,
and is therefore larger in a merger remnant's core than in a single star's core.
In contrast with the first effect, the increased Brunt--V\"ais\"al\"a frequency due to this effect persists until the core heats up enough for the difference in $R_c$ between the merger remnant and single stars to be erased (which occurs on the thermal timescale $\tau_{\mathrm{th,core}}$).
This effect tends to reduce $\Delta\Pi$ in the merger remnant compared to a single RG.

\begin{figure*}
    \centering
    \includegraphics[width=\textwidth]{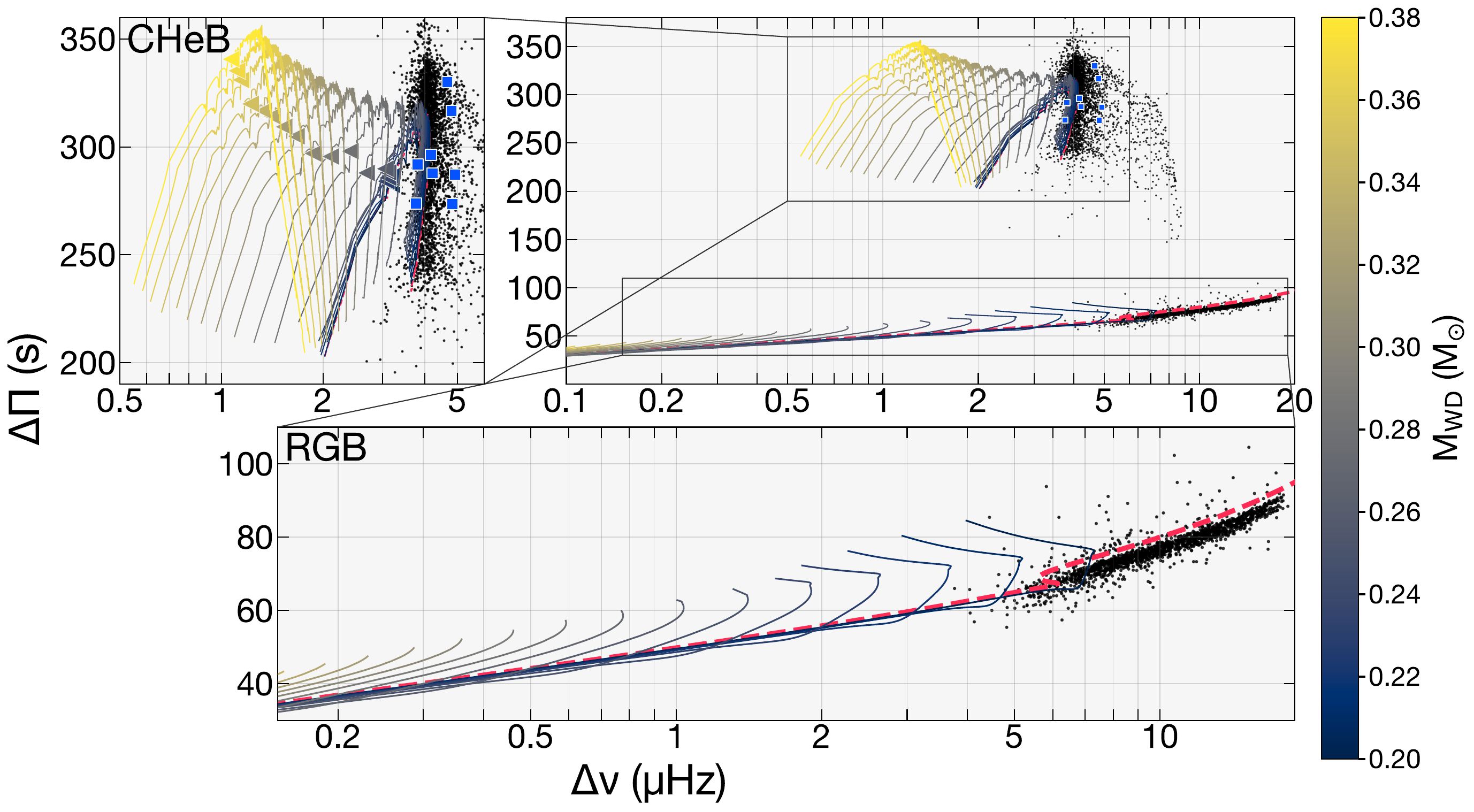}
    \caption{An asteroseismic spacing diagram ($\Delta\Pi$ versus $\Delta\nu$) showing merger remnant evolutionary sequences and a non-merged model (\textit{red dashed}) all with $M=0.80M_\odot$.
    The \textit{top-right} panel shows the entire parameter space, and the \textit{left} and \textit{bottom} panels zoom in on CHeB stars and the RGB, respectively.
    Roughly speaking, both RGB and CHeB stars evolve from right to left on this diagram (i.e., towards decreasing $\Delta\nu$, or increasing $R$).
    The \textit{black points} are observations taken from the catalog of \edit{\citet{vrard2016period} with stars flagged for possible aliases removed}.
    In the \edit{\textit{top-left}} panel, \textit{colored triangles} denote the last $5\,\mathrm{Myr}$ of the CHeB phase.
    \edit{The \textit{blue squares} denote $8$ stars from the very low-mass sample of \citet{li2022discovery} which appear in the Vrard catalog.}
    }
    \label{fig:asterospacings}
\end{figure*}

As a merger remnant evolves, $\Delta\Pi$ typically evolves from being larger than that of a single star (at \textit{fixed} $M_c$) to \newNZR{smaller} (Figure \ref{fig:asterospacings}), although it may not achieve the latter regime by the helium flash if the merger involves a sufficiently massive WD.
\newNZR{In our models, departures of $\Delta\Pi$ from the non-merged case can reach up to $\approx\!10\,\mathrm{s}$ at a given core mass.}
However, we again caution that $M_c$ is not directly known, and diagnosis of a past merger requires another observable, such as the radius or $\Delta\nu$.

Asteroseismic measurements $\Delta\nu$ and $\Delta\Pi$ of RGs are typically represented on a spacing diagram as shown in Figure \ref{fig:asterospacings}.
Most stars on the RGB (zoomed in on the \textit{bottom} panel) cluster around a tight sequence which ultimately arises from the core mass--luminosity relation \citep[e.g.,][]{deheuvels2022seismic}.
This is also essentially the path followed by our $M=0.8M_\odot$ non-merged model, modulo a small order-unity factor owing to a weak dependence on total mass ($\Delta\nu\propto M^{1/2}$).
In this space, it is clear merger remnants usually lie above the degenerate sequence, before slowly evolving back towards it.
The position of merger remnants above that of normal RGB stars is dominated by the larger radii of these objects at the same core mass.
Since $\Delta\nu\propto R^{-3/2}$, this shifts their positions on the diagram to the left.
\newNZR{Merger remnants may also pass slightly below the degenerate sequence during their evolution (due to sufficiently small values of $\Delta\Pi$).
This effect is however subtle.
Stars in this region of $\Delta\nu$--$\Delta\Pi$ space may more naturally be explained by RG--MS mergers \citep{deheuvels2022seismic}, particularly if they lie far below the degenerate sequence.}

\edit{The g-mode period spacing $\Delta\Pi$ is only observable for remnants sufficiently low on the RGB (i.e., below the RGB bump, which is the ``hook'' feature in Figure \ref{fig:asterospacings}).}
\edit{For merger remnants in this regime, departure from the degenerate sequence is most prominent during the initial contraction phase after the formation of the remnant, although RGs which have passed this phase still lie above the degenerate sequence for an additional $\sim20\,\mathrm{Myr}$.}
Hence, this \edit{particular} asteroseismic diagnosis may only be possible for sufficiently recent mergers involving sufficiently low-mass He WDs.


\subsection{Overbright tip of the RGB} \label{trgb}

During the RGB phase of a single star, the helium core grows in mass and contracts over time, as the envelope expands.
Simultaneously, the core heats up gradually until the temperature is high enough for helium burning through the triple-$\alpha$ process ($\simeq10^8\,\mathrm{K}$), at which time the star has reached the tip of the RGB (tRGB).


In our most extreme model with $M_{\mathrm{WD}}=0.38M_\odot$, the tRGB surface luminosity exceeds that of a single star by a factor of $2$ (Figure \ref{fig:hrd_flashclump}).
This factor is likely to be even larger for more massive values of $M_{\mathrm{WD}}$ than we can run (but which are still physical).
At the tRGB, our merger remnant models outshine the tRGB of the single star model for between $140\,\mathrm{kyr}$ ($M_{\mathrm{WD}}=0.20M_\odot$) and $1.8\,\mathrm{Myr}$ ($M_{\mathrm{WD}}=0.38M_\odot$).
\newNZR{Since RG luminosities are essentially totally determined by near-core properties, even merger remnants of low mass (involving small values of $M_{\mathrm{MS}}$) are still expected to outshine single RGs at the tRGB.}

\newNZR{The ostensibly well-known luminosity of the tRGB is often leveraged to measure cosmological distances \citep{bellazzini2001step,bellazzini2004calibration}.
We point out that overbright merger remnants near the tRGB may affect the tRGB of a stellar population's role as a standard candle.
However, this effect is probably minor, since merger remnants of this type likely make up no more than a few percent of all RGs.}


\begin{figure}
    \centering
    \includegraphics[width=0.48\textwidth]{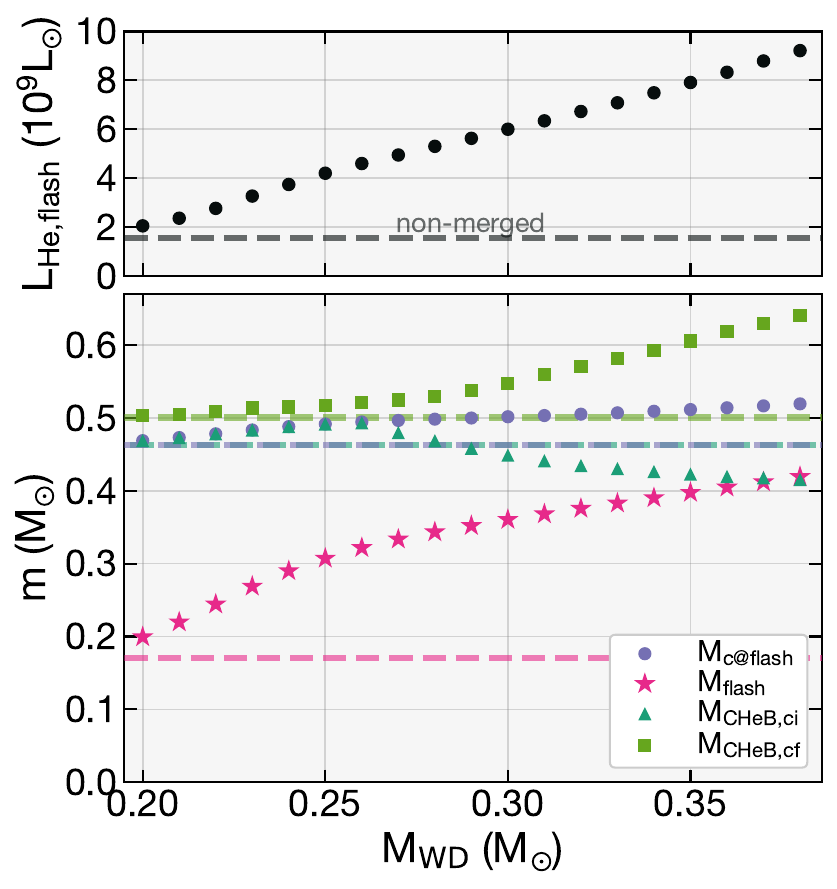}
    \caption{\textit{Top}: The maximum helium-burning luminosities $L_{\mathrm{He,flash}}$ attained during the helium flash in our merger remnant models with total mass $M_{\mathrm{tot}}=M_{\mathrm{MS}}+M_{\mathrm{WD}}=0.80M_\odot$.
    \textit{Bottom:} For the merger remnant models, we show core masses $M_{\mathrm{c@flash}}$ at the helium flash (\textit{purple circles}), mass coordinates $M_{\mathrm{flash}}$ where the helium flash begins (\textit{magenta stars}),
    and initial (\textit{turquoise triangles}) and final (\textit{green squares}) core masses $M_{\mathrm{CHeB,ci}}$ and $M_{\mathrm{CHeB,cf}}$ during the CHeB phase.
    In both panels, \textit{dashed lines} denote values for the $M=0.80M_\odot$ non-merged model.
    }
    \label{fig:helium_flash_v2}
\end{figure}


\section{Core helium-burning phase} \label{chebsupersect}

When the core reaches a sufficient temperature, helium burning begins.
In single RGs, it is well known that helium ignites off-center due to a slight temperature inversion caused by neutrino cooling \citep{thomas1967sternentwicklung}.
The peak burning rate reached during the subsequent helium flash depends sensitively on the density where helium ignites \citep{salpeter1957nuclear}, and therefore on the mass of the helium core.
Subsequent intermittent burning events occur in a series of subflashes \citep{thomas1967sternentwicklung}.
In a normal RG, these subflashes propagate inwards over the course of a few megayears until they reach the core, fully lifting the degeneracy of the helium core and burning $\approx\!4\%$ of its mass into carbon.
Once this occurs, RGs quickly contract on a thermal time until they have radii $\approx\!11R_\odot$ (the CHeB phase), after which they are supported by a combination of helium core- and hydrogen shell burning, which generally contribute in comparable amounts to the stellar luminosity.

\begin{figure*}
    \centering
    \includegraphics[width=\textwidth]{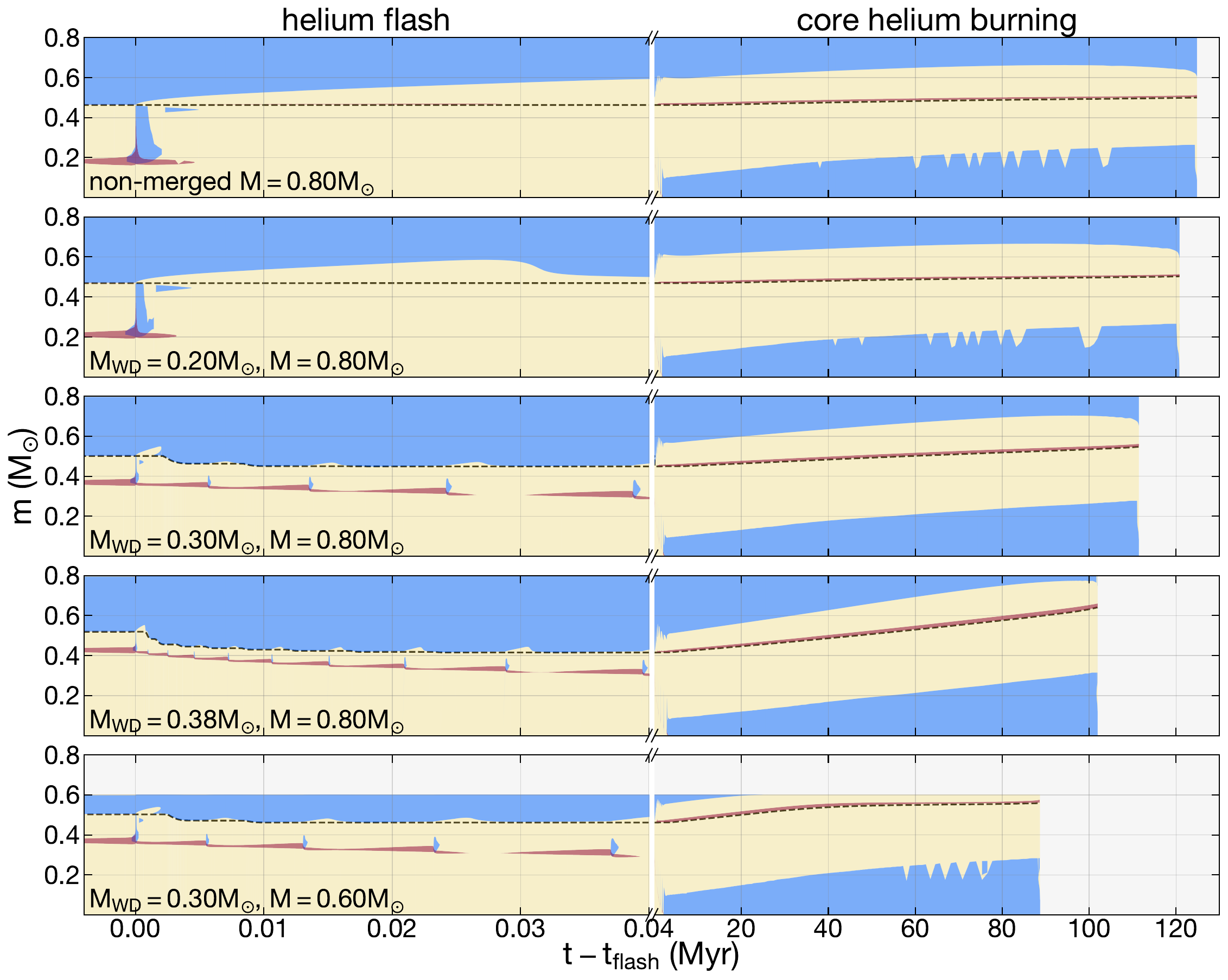}
    \caption{Kippenhahn diagrams for the non-merged $M=0.80M_\odot$ and selected merger remnant models, during the helium flash (\textit{left}) and CHeB phase (\textit{right}).
    Convective (radiative) zones are shown in \textit{blue} (\textit{yellow}), and burning regions fusing more vigorously than $\epsilon=10^3\,\mathrm{erg}\,\mathrm{g}^{-1}\,\mathrm{s}^{-1}$ are shown in \textit{red} (at this scale, the CHeB region is highly localized to the center of the star and cannot be seen).
    The \textit{black dashed line} indicates the helium core mass $M_c$.
    \newNZR{Merger remnant models with $M_{\mathrm{WD}}\geq0.27M_\odot$ experience core dredge-up events during the helium flash, and subsequently develop thinner convective envelopes by mass during CHeB. If $M_{\mathrm{MS}}$ is small enough, the convective envelope disappears entirely, and the star enters the horizontal branch.}}
    \label{fig:a_fleet_of_kippenhahns}
\end{figure*}

In merger remnants, the helium flash occurs similarly, although, as previously pointed out by \citet{zhang2017evolution}, the flash occurs farther off-center (Figure \ref{fig:helium_flash_v2}), since the core temperature inversion is now dominated by the low temperature of the WD progenitor rather than the weaker effect of neutrino cooling (compare the \textit{dashed red} and \textit{solid blue curves} of the \textit{lower-right} panel of Figure \ref{fig:explain_period_spacings}).
Additionally, the subflashes occur closer together in time, and are more energetic in general.

In our models, the maximum helium-burning luminosity attained by a merger remnant is roughly a linear function of $M_{\mathrm{WD}}$, with the most extreme model attaining a factor of $\approx \! 5$ higher helium-burning luminosity than attained by a non-merged star (see the \textit{top} panel of Figure \ref{fig:helium_flash_v2}).
The energy production rate can be a whopping $\approx \! 10^{10} L_{\odot}$ at the peak of the helium flash.
Since mergers involving more massive WDs than those on our model grid are possible, even larger helium-burning luminosities may occur in nature.
The most vigorous helium flash in our merger models also burns a larger fraction of the core into carbon, up to $\simeq7\%$ by mass.
\edit{We present the post-helium flash composition profiles of several of our models in Appendix \ref{chebcomp}.}

We verify in our models that the helium flash, though abnormally energetic in merger remnants, still does not result in hydrodynamical burning (which might result in detonation).
Specifically, following \citet{shen2010thermonuclear}, we verify that convection is always efficient enough to flatten temperature gradients created by helium burning, and also that the hierarchy $|v_r|\ll v_{\mathrm{conv}}\ll c_s$ is maintained throughout the helium core (though not necessarily in the hydrogen envelope).
\newNZR{The properties of this dredge-up concern physics in the helium core only, and is essentially independent of the envelope mass.}
\newNZR{Similar core dredge-up events during the helium flash have also been predicted in low-metallicity stars which have undergone extreme mass loss \citep{sweigart1997helium,cassisi2002first}.}

The CHeB-phase evolution of our merger remnant models fits into two regimes, sorted by $M_{\mathrm{WD}}$:
\begin{enumerate}
    \item For $M_{\mathrm{WD}}\leq0.26M_\odot$, the flash is delayed slightly, and occurs when the core is slightly more massive.
    As a result, the CHeB phase of these remnants begins with a slightly higher helium core mass.
    Specifically, our single non-merged model ignites helium at a core mass $M_c\approx0.46M_\odot$, and our merger remnant model with $M_{\mathrm{WD}}=0.26M_\odot$ undergoes its flash at a slightly higher core mass $M_c\approx0.50M_\odot$.
    \item For $M_{\mathrm{WD}}\geq0.27M_\odot$,
    the flash produces so much heat that it causes the convective envelope to deepen significantly, dredging up both the hydrogen-burning shell and the outer layers of the helium core (see Figure \ref{fig:a_fleet_of_kippenhahns}).
    For our most extreme model ($M_{\mathrm{WD}}=0.38M_\odot$), $\approx\!20\%$ of the core ($\approx0.1M_\odot$ of helium) is dredged up into the envelope, significantly modifying its mean molecular weight.
\end{enumerate}

\begin{figure*}
    \centering
    \includegraphics[width=\textwidth]{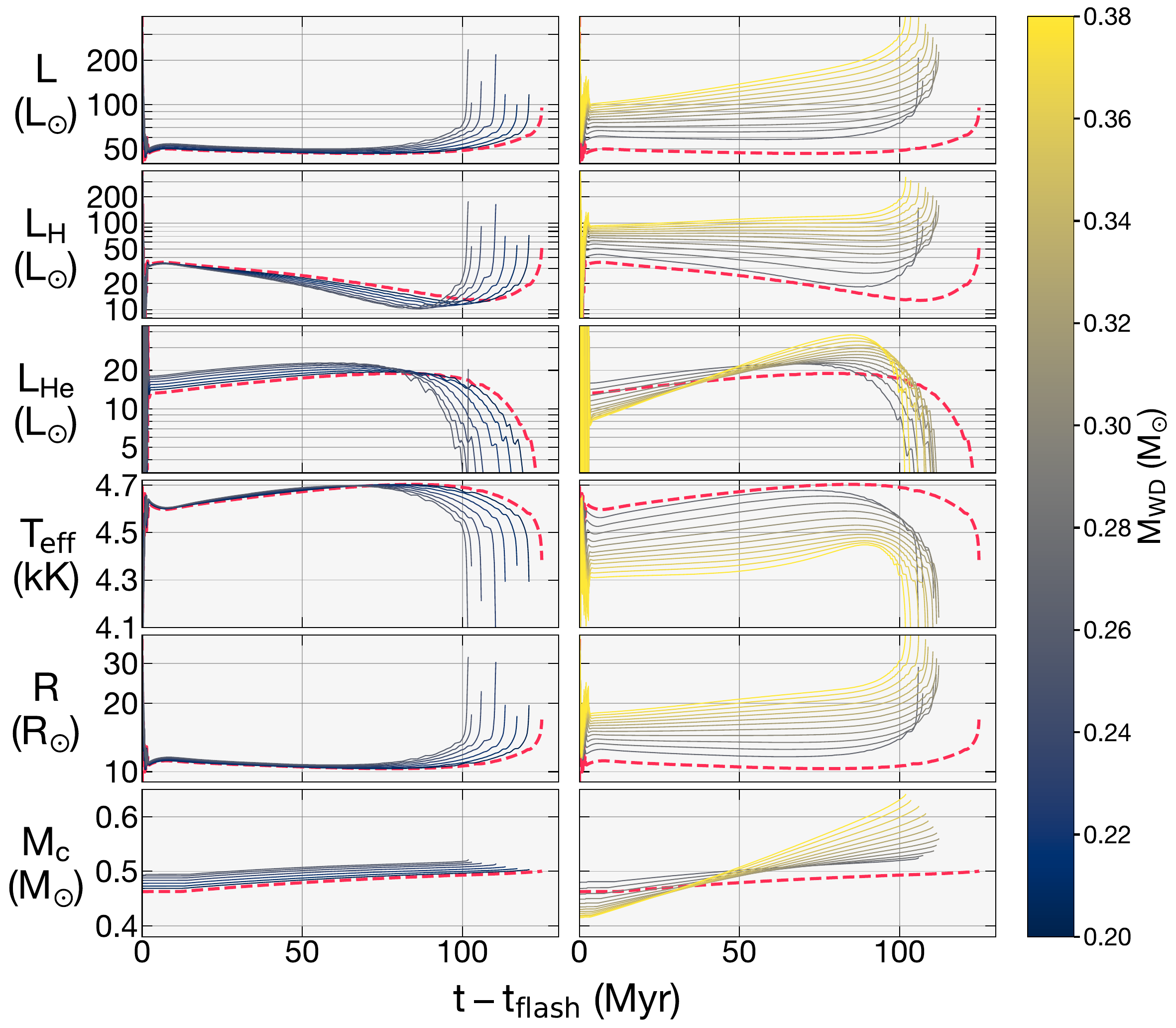}
    \caption{Time evolution of the CHeB-phase luminosity $L$, hydrogen- and helium-burning luminosities $L_{\mathrm{H}}$ and $L_{\mathrm{He}}$, effective temperature $T_{\mathrm{eff}}$, radius $R$, and helium core mass $M_c$ for merger remnant models and a non-merged model (\textit{red dashed}) all with $M=0.80M_\odot$.
    The \textit{left} and \textit{right} panels show merger remnant models with $M_{\mathrm{WD}}\leq0.26M_\odot$ and $M_{\mathrm{WD}}\geq0.27M_\odot$, respectively.
    The latter models have helium flashes which are vigorous enough to dredge up some of the helium core into the envelope.}
    \label{fig:clump_quantities_over_time}
\end{figure*}

As a result of this, merger remnants involving lower-mass He WDs $M_{\mathrm{WD}}\leq0.26M_\odot$ evolve very similarly to normal red clump stars, appearing almost identical to them in both photometric (\textit{bottom-left} panel of Figure \ref{fig:hrd_flashclump}) and asteroseismic (\textit{top-left} panel of Figure \ref{fig:asterospacings} and \textit{left} panel of Figure \ref{fig:clump_quantities_over_time_astero}) observables.
One minor difference is that the merger remnants have systematically shorter CHeB lifetimes, by up to $\lesssim20\%$.

In contrast, merger remnants involving more massive He WDs $M_{\mathrm{WD}}\geq0.27M_\odot$ evolve and appear very differently from their single counterparts.
The following sections focus on observational signatures in this latter case (hereafter \textit{post-dredge-up merger remnants}).
The dredge-up event ultimately affects the luminosity (Section \ref{overluminous}), asteroseismology (Sections \ref{chebdeltapg} and \ref{chebcoupling}), and surface abundances (Section \ref{chebabund}) significantly.

\subsection{Over-luminous red clump stars} \label{overluminous}

During the CHeB phase, post-dredge-up remnants have significantly modified photometric properties:
\begin{enumerate}
    \item They are brighter and slightly cooler, with the effect being stronger for higher values of $M_{\mathrm{WD}}$ (\textit{bottom-left} panel of Figure \ref{fig:hrd_flashclump} and \textit{top-right} panel of Figure \ref{fig:clump_quantities_over_time}).
    For our most extreme model ($M_{\mathrm{WD}}=0.38M_\odot$), the luminosity is roughly tripled.
    \item Their luminosities evolve much more significantly over the CHeB phase.
    Again, this effect is stronger for higher values of $M_{\mathrm{WD}}$.
    In comparison, normal CHeB stars have essentially fixed luminosities for almost the entirety of the helium-burning phase.
\end{enumerate}

As in normal CHeB stars, nuclear energy production has two main contributions:
\begin{enumerate}
    \item \textit{Core helium burning.}
    The now non-degenerate helium core burns helium through the triple-$\alpha$ process in a convective core.
    Structurally, the core is similar to normal, intermediate-mass MS stars in that they have convective core burning surrounded by a radiative envelope.
    Analogously, the helium-burning luminosity is essentially set by the helium core mass.

    \item \textit{Hydrogen shell burning.}
    Hydrogen continues to burn in a shell around the helium core through the CNO cycle.
    The luminosity of the hydrogen burning is a sensitive function of the environment around the burning shell.
\end{enumerate}

The luminosities of these burning regions are highly coupled to each other.
For example, the mass of the helium core determines the \textit{helium-burning} luminosity and, thus, the radius of the core.
This in turn strongly affects the \textit{hydrogen-burning} luminosity and thus the total luminosity of the star.
In turn, the luminosity of the hydrogen-burning shell determines the growth rate of the helium core, therefore feeding back onto the time dependence of \textit{both} contributions to the luminosity.

As can be seen in the \textit{right} panels of Figure \ref{fig:clump_quantities_over_time}, post-dredge-up merger remnants can have significantly larger hydrogen-burning luminosities $L_{\mathrm{H}}$.
While normal clump stars have comparable hydrogen- and helium-burning luminosities ($\approx \! 20L_\odot$ in both cases), our most extreme merger remnant model with $M_{\mathrm{WD}}=0.38M_\odot$ has $L_{\mathrm{H}}\simeq100L_\odot$ (compared to $L_{\mathrm{He}}\lesssim40L_\odot$).

The dominant factor setting $L_{\mathrm{H}}$ in a post-dredge-up remnant is the mean molecular weight $\mu$ at the hydrogen-burning shell, which is significantly enhanced by the helium dredge-up event.
Single star models have a near-solar helium mass fraction $Y\approx30\%$.
In contrast, the $M_{\mathrm{WD}}=0.38M_\odot$ remnant model (with total mass $M=0.80M_\odot$) has a much larger value $Y\approx46\%$ owing to the $\approx \! 0.1M_\odot$ of helium added to a pre-existing, solar-composition envelope with mass $\approx0.3M_\odot$.
These values of $Y$ correspond approximately to mean molecular weights $\mu\approx0.62$ and $\mu\approx0.70$, respectively.
\citet{refsdal1970core} show that the hydrogen-burning luminosity exhibits a steep scaling with $\mu$: $L\propto\mu^{7\text{--}8}$ for CNO-cycle burning,
corresponding to increases in $L_{\mathrm{H}}$ by a factor $\simeq2.5$, which is roughly consistent with the behavior of $L_{\mathrm{H}}$ in our models.

Because $L_{\mathrm{H}}$
is very sensitive to $\mu$,
the appearance and evolution of the remnant is now acutely sensitive to the initial envelope mass of the remnant.
This is contrary to single stars, in which $L$ and $T_{\mathrm{eff}}$ during the CHeB phase are nearly independent of the envelope mass.
Since $\mu$ is set by the final mass fraction of the envelope after the dredge-up event, 
larger pre-existing envelopes dilute the added helium and reduce the mean molecular weight enhancement.
We note in passing that structurally important increases in $\mu$ during a core dredge-up event also appear in some RG--He WD merger models of \citet{zhang2013white}.

As can be seen in Figure \ref{fig:clump_quantities_over_time}, the helium-burning luminosity $L_{\mathrm{He}}$ is also different in post-dredge-up models.
$L_{\mathrm{He}}$ essentially tracks the helium core mass $M_c$ until significant core helium depletion at the end of the CHeB phase.
The high value of $L_{\mathrm{H}}$ translates to a fast-growing helium core, and, in turn, an up to a factor of a few increase in $L_{\mathrm{He}}$ over the CHeB phase.
In sum, the higher $L_{\mathrm{H}}$ and $L_{\mathrm{He}}$ naturally translate to an overbright CHeB phase with significant time evolution in luminosity.

The increased luminosity
causes merger remnants to have abnormally large radii,
up to $R\approx20R_\odot$ in our models (for $M_{\mathrm{WD}}=0.38M_\odot$).
In addition to affecting the photometry, these large radii can also be directly measured asteroseismically via $\nu_{\mathrm{max}}$ and $\Delta\nu$, together with the scaling relations in Equations \ref{numax} and \ref{dnu}.
However, in order to distinguish merger remnants from ordinary stars beginning to ascend the asymptotic giant branch, it may also be necessary to measure mixed mode period spacings or surface abundances.

\subsection{Wider range of asteroseismic g-mode period spacings} \label{chebdeltapg}

\begin{figure*}
    \centering
    \includegraphics[width=\textwidth]{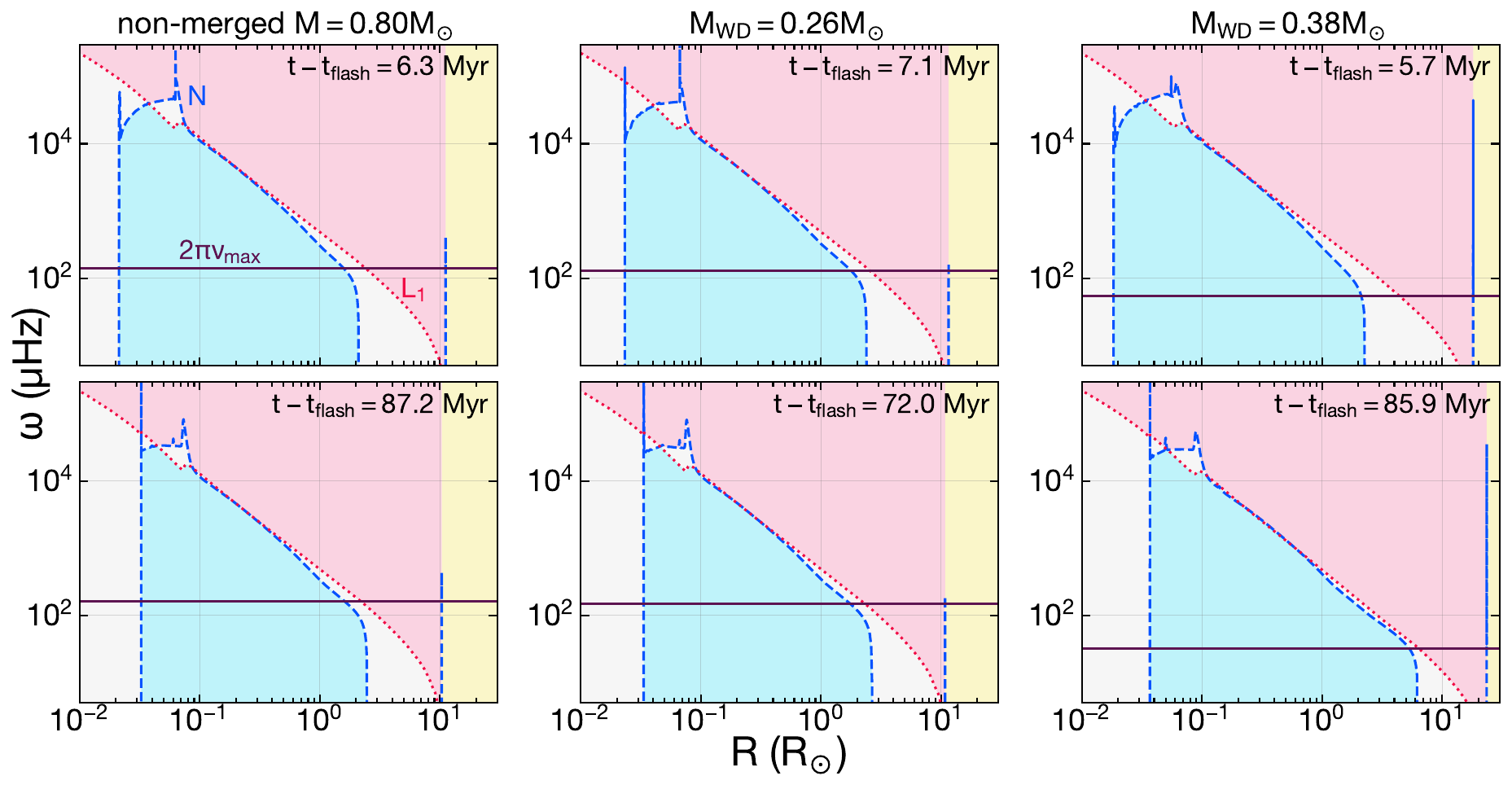}
    \caption{CHeB-phase asteroseismic propagation diagrams for a non-merged model (\textit{left}) and merger remnants with $M_{\mathrm{WD}}=0.26M_\odot$ (\textit{center}) and $M_{\mathrm{WD}}=0.38M_\odot$ (\textit{right}).
    The Brunt--V\"ais\"al\"a ($N$), dipole Lamb ($L_1$), and maximum power ($2\pi\nu_{\mathrm{max}}$) frequencies are shown as the \textit{dashed blue}, \textit{dotted red}, and \textit{solid purple} curves, respectively.
    Color coding of areas denotes the p-mode propagating regions (\textit{pink}), g-mode propagating regions (\textit{light blue}), evanescent regions (\textit{white}), and the exterior of the star (\textit{yellow}).
    \textit{Top} panels show propagation diagrams near the beginning of the CHeB phase, and \textit{bottom} panels
    show the same farther along the CHeB phase, when the models attain maxima in the asteroseismic g-mode period spacing $\Delta\Pi$.
    All models have total mass $M=0.80M_\odot$.}
    \label{fig:merger_propagation_diagrams}
\end{figure*}

\begin{figure*}
    \centering
    \includegraphics[width=\textwidth]{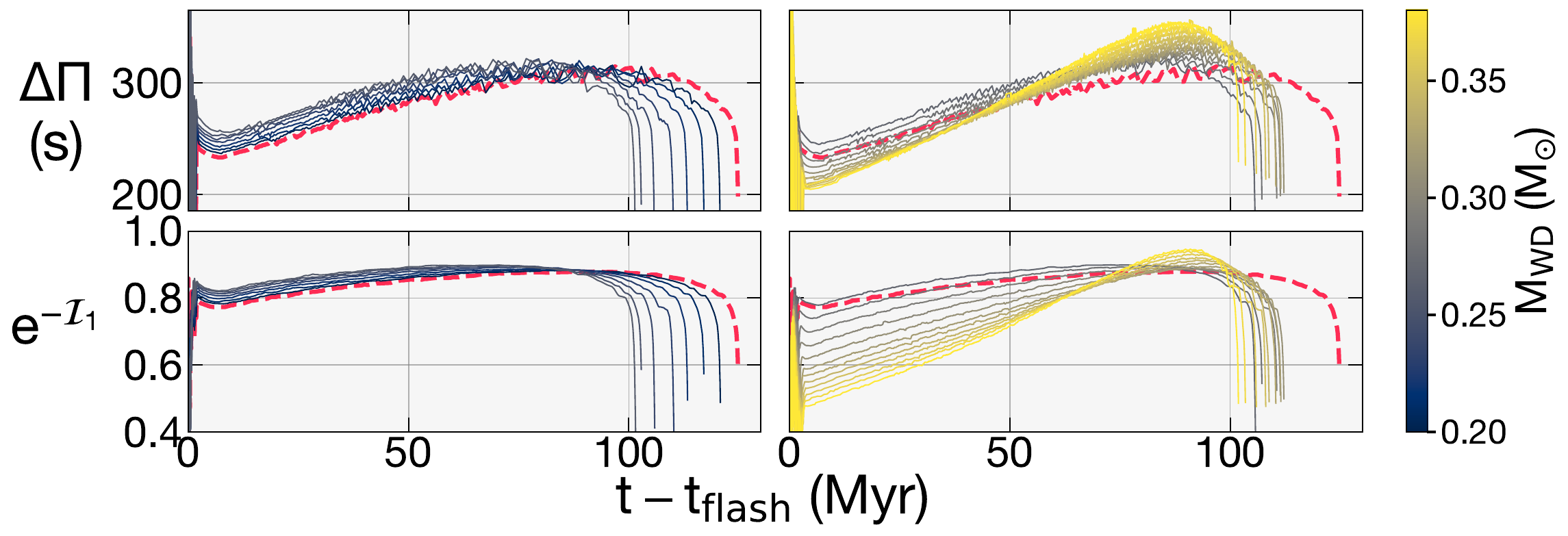}
    \caption{Time evolution of the CHeB-phase asteroseismic g-mode period spacing $\Delta\Pi$ (\textit{top}), and $e^{-\mathcal{I}_1}$ (\textit{bottom}), which is a simple proxy for the dipole mixed-mode coupling factor. We plot merger remnant models and a non-merged model (\textit{red dashed}) all with $M=0.80M_\odot$.
    As in Figure \ref{fig:clump_quantities_over_time}, the \textit{left} and \textit{right} panels show merger remnant models with $M_{\mathrm{WD}}\leq0.26M_\odot$ and $M_{\mathrm{WD}}\geq0.27M_\odot$, respectively.}
    \label{fig:clump_quantities_over_time_astero}
\end{figure*}

During the CHeB phase, post-dredge-up mergers remnant models have substantially different structures which directly modify asteroseismic observables such as $\Delta\Pi$.
This can be seen by comparing the \textit{left} and \textit{right} panels of Figure \ref{fig:merger_propagation_diagrams}, which show the propagation diagrams of the non-merged and $M_{\mathrm{WD}}=0.38M_\odot$ merger remnant models at two selected times.
On the other hand, merger remnants which do not experience significant dredge-up of the helium core ($M_{\mathrm{WD}}\leq0.26M_\odot$) have essentially identical structures (compare the \textit{left} and \textit{center} panels of Figure \ref{fig:merger_propagation_diagrams}).

Because the nuclear burning luminosities are substantially modified in post-dredge-up models, the g-mode period spacing $\Delta\Pi$ of merger remnants spans a wider range of values over the CHeB phase (see the \textit{top-right} panel of Figure \ref{fig:clump_quantities_over_time_astero}, as well as the \textit{top-left} panel of Figure \ref{fig:asterospacings}).
While $\Delta\Pi$ (Equation \ref{dpgformula}) depends on the integral of $N$ within the radiative core,
its value is dominated by the regions with the largest $N$.
As can be seen in Figure \ref{fig:merger_propagation_diagrams}, $N$ is maximal underneath the hydrogen-burning shell (i.e., below the compositional spike in $N$ at radii $\lesssim0.1R_\odot$).

At the beginning of the CHeB phase (the \textit{top} panels of Figure \ref{fig:merger_propagation_diagrams}), post-dredge-up merger remnants have smaller helium core masses, resulting in smaller $L_{\mathrm{He}}$.
This results in a smaller convective core (slightly extending the bottom boundary of the g-mode cavity) as well as a slightly higher core dynamical frequency (resulting in slightly higher values of $N$ overall).
Both of these effects tend to decrease $\Delta\Pi$.

In contrast, near the end of CHeB (the \textit{bottom} panels of Figure \ref{fig:merger_propagation_diagrams}), the helium core of merger remnants has grown substantially due to the large hydrogen burning rate.
They therefore evolve to have larger $L_{\mathrm{He}}$, which ultimately results in an \textit{increased} convective core size and overall lower $N$ profile, and therefore larger $\Delta\Pi$.
Because they also have inflated radii (Section \ref{overluminous}), they have smaller values of $\Delta\nu$, and will occupy a region to the left of normal clump stars on a $\Delta\nu$--$\Delta\Pi$ spacing diagram (\textit{top} panels of Figure \ref{fig:asterospacings}).

\subsection{Asteroseismic mixed-mode coupling} \label{chebcoupling}

In RGs, pulsations can probe not only the g-mode cavity (through $\Delta\Pi$) but also the evanescent zone between the p- and g-mode cavities.
The extent of coupling between the two cavities is usually described by a coupling factor $q$ \citep{unno1989nonradial}.
In addition to determining the visibility of mixed modes, $q$ itself is an independent observable which probes a different internal structural feature than does $\Delta\Pi$ \citep{mosser2017period,dhanpal2023inferring}.
The recent discovery that low-mass or low-metallicity RGs have preferentially high $q$ \citep{matteuzzi2023red,kuszlewicz2023mixed} has created revitalized demand for physical interpretations of $q$.

Computing $q$ is mathematically nontrivial and involves detailed solution of a wave transmission problem \citep[e.g.,][]{takata2016asymptotic,takata2016physical,takata2018asymptotic} or fitting a model spectrum directly \citep[e.g.,][]{jiang2014verification}.
\citet{takata2016asymptotic} show that the coupling factor is related to the transmission coefficient $T$ by
\begin{equation}
    q = \frac{1-\sqrt{1-T^2}}{1+\sqrt{1-T^2}}\mathrm{.}
\end{equation}

\citet{takata2016physical} further write the transmission coefficient as
\begin{equation}
    T = e^{-\pi(X_I+X_R)}\mathrm{.}
\end{equation}

Here, $X_I$ is defined as an integral over the evanescent zone $\mathcal{E}$ of the radial wavenumber with respect to the asymptotic dispersion relation, i.e., for dipole modes,
\begin{equation} \label{xi}
    \begin{split}
        \mathcal{I}_1 \equiv \pi X_I &= \int_{\mathcal{E}}|k_r|\,\mathrm{d}r \\
        &= \int_{\mathcal{E}}\frac{\omega}{c_s}\sqrt{\left(1-N^2/\omega^2\right)\left(L_1^2/\omega^2-1\right)}\,\mathrm{d}r
    \end{split}
\end{equation}

\noindent where, for simplicity, we have applied the Cowling approximation.
$X_R$ is a remainder term 
which can be specified analytically in the limit of a thin evanescent zone \citep{takata2016physical}, where it is most important.

We focus on the value of $e^{-\mathcal{I}_1}$ as a rough proxy for the mixed-mode coupling.
Although CHeB stars typically have strong coupling such that $X_R$ is likely to be important, calculation of this contribution requires more care \citep[see, e.g.,][]{van2023asteroseismology}, and we therefore defer a detailed calculation of $q$ in these merger remnants to a potential future investigation.

From the propagation diagrams (Figure \ref{fig:merger_propagation_diagrams}), it can be seen that the evanescent zone evaluated at $\nu_{\mathrm{max}}$ is initially substantially wider in the $M_{\mathrm{WD}}=0.38M_\odot$ post-dredge-up merger remnant than in the non-merged model.
This decreases the value of $e^{-\mathcal{I}_1}$ and, in turn, the g-mode coupling factor $q$ at early phases of CHeB evolution.

However, late in the CHeB phase of post-dredge-up merger remnants involving especially massive He WDs with $M_{\mathrm{WD}}\geq0.33M_\odot$, the evanescent region becomes exceedingly small because of a larger radiative core.
This temporarily causes $e^{-\mathcal{I}_1}$ and the mixed-mode coupling to become larger than in typical CHeB stars, giving their mixed modes high visibility.

In the last $\lesssim20\,\mathrm{Myr}$, both $\Delta\Pi$ and $e^{-\mathcal{I}_1}$ dive sharply as the CHeB star begins to enter the asymptotic giant branch phase.
During this time, the star expands especially quickly, and it will lie even farther to the left of most merger remnants on the asteroseismic spacing diagram in Figure \ref{fig:asterospacings}.

\begin{figure*}
    \centering
    \includegraphics[width=\textwidth]{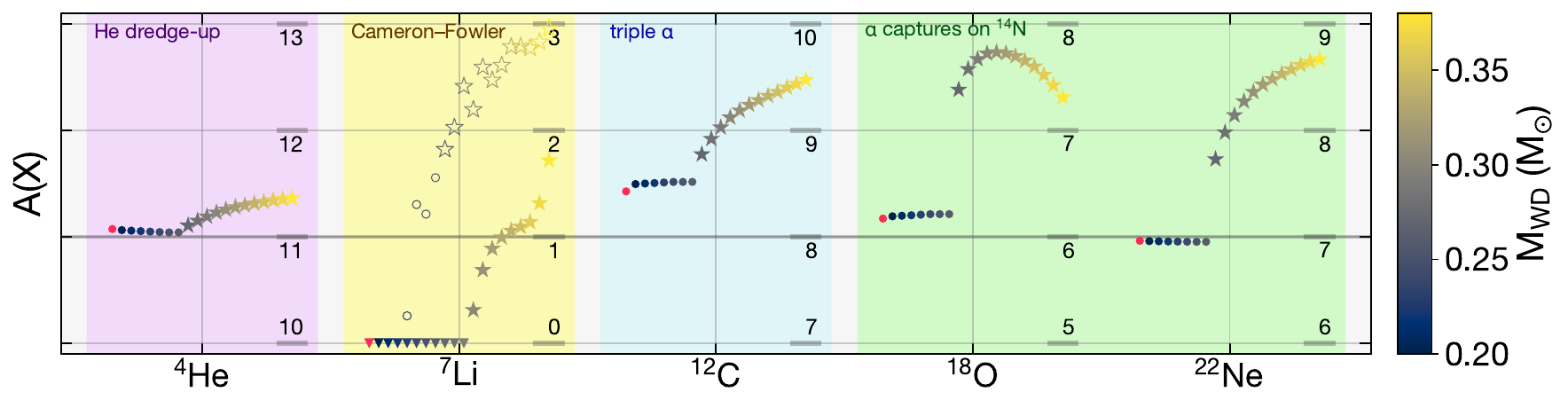}
    \caption{
    Selected surface abundances $\mathrm{A}(\mathrm{X})$ (defined in Equation \ref{ax}) during the CHeB phase.
    \textit{Red points} denote the $M=0.80M_\odot$ non-merged model, and \textit{starred points} denote post-dredge-up merger remnants.
    For ${}^7\mathrm{Li}$, \textit{filled} symbols denote lithium enrichment due to the helium flash only (described in Section \ref{chebabund}), whereas \textit{hollow} symbols denote lithium enrichment due to both the helium flash and merger itself (which may be inaccurately modelled).
    \edit{Note that the vast majority of CHeB stars have $A(\mathrm{Li})\lesssim1.5$ \citep{kumar2020discovery}, and the solar abundance is $A(\mathrm{Li})\approx1.05$ \citep{asplund2009chemical}.}}
    \label{fig:composition_grand_finale}
\end{figure*}

\subsection{Abundance anomalies} \label{chebabund}

Merger remnants that experience a core dredge-up at the flash naturally display unusual surface abundances.
While dredge-up of the helium core most obviously produces an enhancement in ${}^4\mathrm{He}$, direct spectroscopic measurement of helium abundance in late-type stars is likely infeasible.
Fortunately, dredge-up probably enriches the surface with other elements that can be directly probed spectroscopically.
Figure \ref{fig:composition_grand_finale} shows surface abundances for selected species in terms of the index $\mathrm{A}(\mathrm{X})$, defined for a given species $\mathrm{X}$ as
\begin{equation} \label{ax}
    \mathrm{A}(\mathrm{X}) \equiv \log(n_{\mathrm{X}}/n_{\mathrm{H}}) + 12
\end{equation}

\noindent where $n_{\mathrm{X}}$ and $n_{\mathrm{H}}$ are the surface \textit{number} densities of $\mathrm{X}$ and hydrogen, respectively.


We find that post-dredge-up remnants exhibit significant ${}^{12}$C surface enrichment, up to $\simeq1\,\mathrm{dex}$ \edit{relative to hydrogen}.
This enhancement arises from dredge-up of core material which has been partially fused during the helium flash into carbon.
Post-dredge-up remnants thus possess increased values of $\mathrm{C}/\mathrm{Fe}$ and decreased values of ${}^{13}\mathrm{C}/{}^{12}\mathrm{C}$.
Additionally, our post-dredge-up remnant models also possess significant surface enhancements \edit{(up to $\sim1.5\,\mathrm{dex}$ relative to hydrogen)} in ${}^{18}\mathrm{O}$ and ${}^{22}\mathrm{Ne}$, which are created by successive $\alpha$ captures of ${}^{14}\mathrm{N}$ \citep[e.g.,][]{clayton2003handbook} during the helium flash.
\edit{As the surface abundance of $^{16}\mathrm{O}$ in all of our merger remnant models is almost identical to that of our non-merged model, the surface abundance ratio $^{16}\mathrm{O}/^{18}\mathrm{O}$ is also decreased by up to $\sim1.5\,\mathrm{dex}$ in post-dredge-up remnants.}
While not included in our reaction network, it is also probable that a significant amount of ${}^{26}\mathrm{Mg}$ is formed through $\alpha$ capture of ${}^{22}\mathrm{Ne}$ in the abnormally hot helium flash \citep[as in][]{shen2023q}.
The surface abundance of ${}^{26}\mathrm{Mg}$ is therefore also likely to be enhanced.



The lithium abundance is also important, because it can be created by the burning of $^3$He or destroyed by burning at temperatures comparable to those required for hydrogen burning.
Lithium-rich giants \edit{make up about $3\%$ of all CHeB stars \citep{kumar2020discovery}, and} have previously been suggested to have formed via binary interactions or mergers \citep[e.g.,][]{zhang2013white,casey2019tidal}.
\edit{Our} merger remnant models \edit{with high $M_{\mathrm{WD}}$} become highly lithium-rich very soon after merger, owing to a brief dredge-up event which occurs when hydrogen burning is first turned on \citep[similar to the Cameron--Fowler mechanism;][]{cameron1971lithium}.
While such a merger-era lithium enhancement seems \textit{plausible} \citep[see, e.g., observations of lithium enrichment in luminous red novae;][]{kaminski2023lithium}, our treatment of the merger process itself is highly artificial, and the evolution of the remnant is unreliable for post-merger ages \edit{younger} than a thermal time, $\tau_{\mathrm{th,env}}$.
Modification to the surface abundances of other species during this (possibly artificial) dredge-up event are minor, since they primordially occur in much higher abundance than lithium.

If lithium is not enhanced during the merger, we investigate whether core dredge-up during the helium flash may be able to create lithium-rich giants anyway.
To test this, we remove all of the lithium from our merger remnant models (with total $M=0.8M_\odot$) just prior to the tRGB, and evolve them through the helium flash and CHeB phase.
As shown in Figure \ref{fig:composition_grand_finale}, most models do not become lithium-rich according to the standard criterion $\mathrm{A}(\mathrm{Li})\geq1.5$ \citep[e.g.,][]{deepak2021lithium}, with the exception of the $M_{\mathrm{WD}}=0.38M_\odot$ model.
It is possible that merger remnants with higher values of $M_{\mathrm{WD}}$ or lower values of $M_{\mathrm{MS}}$ may attain stronger lithium enhancements.
\newNZR{Moreover, extra mixing during the helium flash suggested by recent evidence may increase the amount of lithium surfaced \citep{kumar2020discovery,martell2021galah,schwab2020helium}.}
Overall, we conclude that, unless lithium enrichment occurs at merger \newNZR{or non-canonical mixing processes operate during the helium flash}, He WD--MS merger remnants are unlikely to become lithium-rich.

\subsection{Populating the horizontal branch with merger remnants} \label{horizontalbranch}

Neglecting mass loss during merger or through winds, $M_{\mathrm{MS}}$ sets the envelope mass of the merger remnant.
In the preceding discussion, we have focused on varying $M_{\mathrm{WD}}$ and fixed $M_{\mathrm{MS}}$ so that the total mass of the merger remnant models is $M_{\rm tot}=M_{\mathrm{WD}}+M_{\mathrm{MS}}=0.80M_\odot$.
However, in principle, $M_{\mathrm{MS}}$ could span a wide range of masses, including very small ones (if the hydrogen-rich component is a brown dwarf or if there is significant mass loss, e.g., \citealt{metzger2021transients}), up to a few solar masses for mergers with intermediate-mass MS stars.
In this section, we explore the behavior of merger remnants under varying $M_{\mathrm{MS}}$.

Variation of $M_{\mathrm{MS}}$ may significantly change the behavior of the remnant in the following ways:
\begin{enumerate}
    \item If $M_{\mathrm{MS}}$ is sufficiently small, merger remnants may only ignite helium with scant hydrogen envelopes (starting CHeB as subdwarf B-type stars; sdBs), or may fail to ignite helium altogether (fizzling out into He WDs).
    For reasons of scope, we do not investigate the He WD or sdB outcomes (but see the detailed \edit{modeling} of \citealt{zhang2017evolution} and \citealt{zhang2023formation}).
    Alternatively, if the envelope mass drops below $M_{\mathrm{env}}\simeq0.1M_\odot$ during CHeB but can still sustain hydrogen shell burning, the remnant can evolve onto the horizontal branch \citep{catelan2009horizontal}.
    \item In post-dredge-up remnants (where a helium flash mixes a fixed amount of core helium into the envelope), the helium fraction of the envelope during the CHeB phase is set by $M_{\mathrm{MS}}$ (larger hydrogen-rich envelopes during this stage more effectively dilute this additional helium).
    For merger remnants massive enough to enter a CHeB phase with a hydrogen-burning shell, this significantly affects their hydrogen-burning (and, thus, total) luminosity (as in Section \ref{chebsupersect}).
\end{enumerate}

\begin{figure*}
    \centering
    \includegraphics[width=\textwidth]{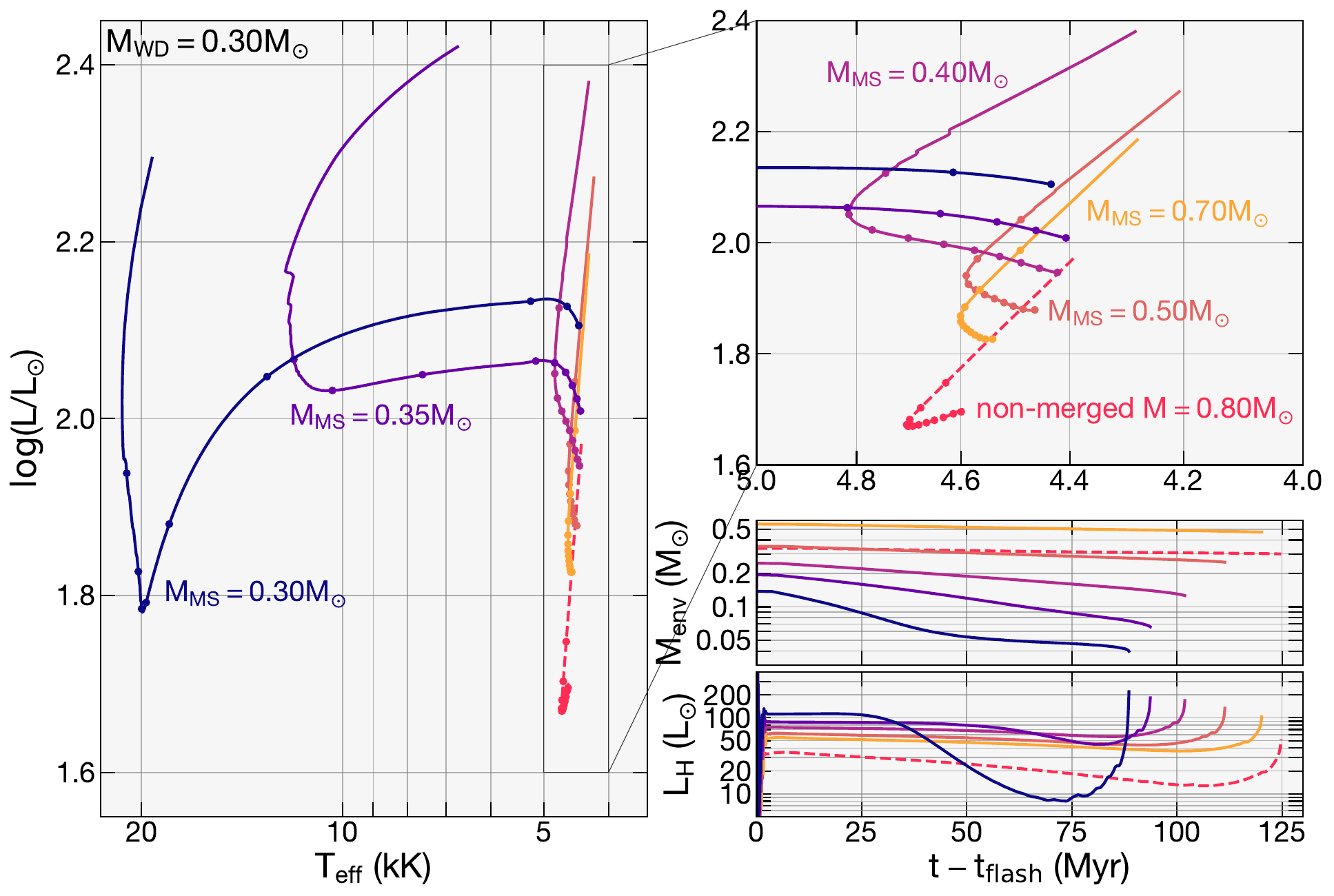}
    \caption{Evolution during the CHeB phase for merger remnants with $M_{\mathrm{WD}}=0.30M_\odot$ but varying $M_{\mathrm{MS}}$ (Section \ref{horizontalbranch}).
    We plot a Hertzsprung--Russell diagram (\textit{left}, with zoomed-in inset on the \textit{top right}) and the time evolution of the hydrogen-rich mass $M_{\mathrm{env}}=M-M_c$ and hydrogen shell-burning luminosity $L_{\mathrm{H}}$ (\textit{bottom right}).
    A non-merged model with $M=0.80M_\odot$ is also shown for comparison (\textit{red dashed line}).
    }
    \label{fig:weird_horizontal_branch}
\end{figure*}

Fixing $M_{\mathrm{WD}}=0.30M_\odot$, we present in Figure \ref{fig:weird_horizontal_branch} the CHeB-phase evolution of merger remnants with $M_{\mathrm{MS}}/M_\odot\in[0.30,0.35,0.40,0.50,0.70]$. 
All of these models are massive enough to reach the tRGB and undergo a helium flash, which in this case is energetic enough to dredge up $\approx0.06M_\odot$ of helium.
However, the envelopes of models with lower $M_{\mathrm{MS}}$ possess much more helium-enriched envelopes: envelope helium mass fractions during the CHeB phase for these models range from $Y=0.36$ (for $M_{\mathrm{MS}}=0.70M_\odot$) to $Y=0.48$ (for $M_{\mathrm{MS}}=0.30M_\odot$).

At the zero-age CHeB, all of these models possess a convective envelope.
As expected, models with lower $M_{\mathrm{MS}}$ (and higher envelope $Y$) have higher hydrogen shell-burning luminosities at the beginning of the CHeB phase (as can be seen in the $L_{\mathrm{H}}$ panel in Figure \ref{fig:weird_horizontal_branch}).
Models with higher $M_{\mathrm{MS}}=0.40$, $0.50$, and $0.70M_\odot$ retain these convective envelopes and behave similarly to the remnants discussed in Section \ref{chebsupersect}.

The lower-mass $M_{\mathrm{MS}}=0.30$ and $0.35M_\odot$ models display significantly different behavior.
These models burn most of their remaining hydrogen during the CHeB phase such that their envelope mass drops below $0.1 \, M_\odot$.
The outer layers of these models become completely radiative (at $t-t_{\mathrm{flash}}\approx25\,\mathrm{Myr}$ and $55\,\mathrm{Myr}$, respectively), and the remnants behave like horizontal branch stars.
When this occurs, the stars become very blue, roughly reaching respective effective temperatures $T_{\mathrm{eff}}\approx20000\,\mathrm{K}$ and $\approx12000\,\mathrm{K}$.
While these models continue to sustain hydrogen-shell burning to some extent, $L_{\mathrm{H}}$ significantly drops during this horizontal branch stage (decreasing by factors $\simeq10$ and $\gtrsim2$ for the $M_{\mathrm{MS}}=0.30M_\odot$ and $M_{\mathrm{MS}}=0.35M_\odot$ models).
In the $M_{\mathrm{MS}}=0.30M_\odot$ model, this extreme drop in $L_{\mathrm{H}}$ precipitates a significantly lower total luminosity, which is readily apparent on a Hertzsprung--Russell diagram (\textit{top-left} panel of Figure \ref{fig:weird_horizontal_branch}).
We confirm for these two cases that the inclusion of gravitational settling does not change the results.
\newNZR{A Reimers wind \citep{reimers1975circumstellar,reimers1977absolute} scaled as in \citet{reimers1977absolute} suggests that winds in these objects are small ($\dot{M}<10^{-10}\,M_\odot\,\mathrm{yr}^{-1}$) and may be ignored.}


\section{Candidate merger remnants} \label{candidates}

\subsection{Undermassive red clump stars may be merger remnants}

\begin{figure}
    \centering
    \includegraphics[width=0.46\textwidth]{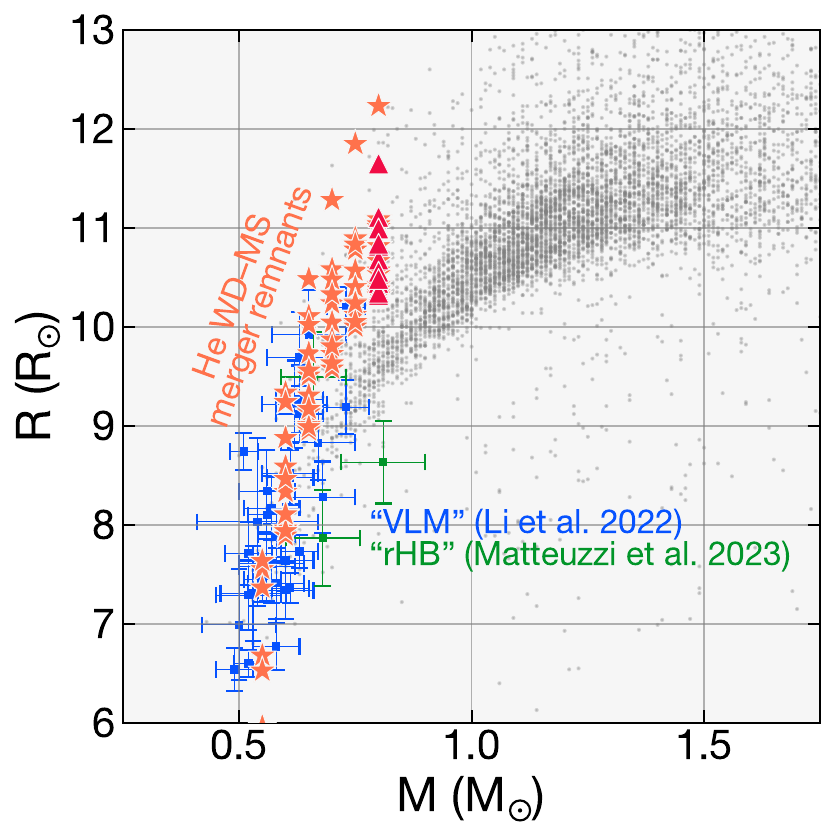}
    \caption{Measured masses and radii of observed CHeB stars, measured using asteroseismology.
    \textit{Blue} and \textit{green squares} denote the \textit{very low-mass} (VLM) sample of \citet{li2022discovery} and \textit{red horizontal branch} (rHB) stars of \citet{matteuzzi2023red}.
    \textit{Gray data points} are taken from the catalog of \citet{yu2018asteroseismology}.
    \textit{Orange stars} denote CHeB-phase merger remnant models with $M_{\mathrm{WD}}=0.20M_\odot$ and varying $M_{\mathrm{MS}}$.
    \textit{Red triangles} denote a CHeB-phase non-merged model with $M=0.8M_\odot$, for comparison.
    Model track points are sampled $10\,\mathrm{Myr}$ apart.}.
    \label{fig:li_undermassive_clump}
\end{figure}

Using asteroseismology, \citet{li2022discovery} discovered a population of undermassive CHeB stars with masses $\lesssim0.8M_\odot$ (their ``very low-mass'' sample).
Since single stars of these masses could not have evolved off of the MS in the age of the universe, these undermassive stars must have undergone non-standard evolution, such as stripping by a companion or binary assembly.
\newNZR{Although \citet{li2022discovery} argue that these undermassive giants are the product of partial envelope stripping by close companions, most of these objects do not exhibit the expected radial velocity variability between APOGEE and Gaia (Kareem El-Badry, private communication), disfavoring this formation channel.}
\citet{matteuzzi2023red} identify several more such objects (referred to as ``red horizontal branch'' stars), further demonstrating their extremely strong mixed-mode coupling.
\edit{Figure \ref{fig:asterospacings} shows $8$ members of the very low-mass sample of \citet{li2022discovery} for which \citet{vrard2016period} reports g-mode period spacings.
Despite their small masses, these very low-mass CHeB stars have typical values of $\Delta\Pi$ and, therefore, likely possess similar core structures to those of normal CHeB stars.}

Observed undermassive clump stars likely have typical helium core masses and only stand out due to their low envelope masses.
He WD--MS mergers naturally explain these objects: as long as $M_{\mathrm{WD}}$ is small enough that no dredge-up occurs at the helium flash, merger remnants have core masses which are basically normal.
Furthermore, a merger remnant's envelope mass is simply set by $M_{\mathrm{MS}}$ (modulo merger or tRGB mass loss), which can be arbitrarily low.
Finally, this binary scenario does not leave any companion behind, explaining why most undermassive CHeB stars are consistent with being single at present.
While other mechanisms may exist for forming single undermassive CHeB stars (e.g., mass loss during failed common-envelope events), He WD--MS mergers appear to be highly promising.

To demonstrate this possibility, we run additional merger remnant models, fixing $M_{\mathrm{WD}}=0.20M_\odot$ and varying $M_{\mathrm{MS}}$ between $0.35M_\odot$ and $0.60M_\odot$ (total masses $M$ between $0.55M_\odot$ and $0.80M_\odot$).
Lower-mass merger remnant models have masses and radii which are consistent with the observed very low-mass sample of \citet{li2022discovery} (Figure \ref{fig:li_undermassive_clump}).
Because the He WD--MS merger channel produces CHeB stars with essentially normal cores, our models behave almost identically to models performed by \citet{li2022discovery} of normal CHeB stars with artificial envelope stripping to mimic mass loss from an initially standard RG.

\subsection{Zvrk: a possible post-dredge-up merger remnant?}

Using asteroseismology, \citet{ong2024gasing} recently discovered a peculiar RG (``Zvrk'') with the following features:
\begin{enumerate}
    \item The oscillation spectrum is complex, superficially resembling typical spectra of CHeB giants \citep[which are dense due to their strong mixed-mode coupling, e.g.,][]{mosser2017period,dhanpal2023inferring}.
    
    \item The asteroseismic scaling relations imply a radius $R\approx24R_\odot$, a factor of two larger than that of a typical CHeB star.
    
    \item The star is highly lithium-rich ($\mathrm{A(Li)}>3$), and also \edit{has a somewhat high [C/N] ratio} relative to typical first-ascent RGs \edit{of the same mass}, or indeed CHeB stars \citep[e.g.,][]{bufanda2023investigating}.

    \item Photometric modulation and asteroseismic rotational splittings (assuming a pure p-mode spectrum) are consistent with a fast-spinning envelope with period $P\approx100\,\mathrm{d}$.
\end{enumerate}

Mainly on the basis of its radius, \citet{ong2024gasing} conclude that Zvrk cannot be in the CHeB phase and must instead be a first-ascent RG.
Because the expected mixed-mode coupling for an RG of this size would be weak, they argue that the spectrum is actually composed of pure p modes (rather than mixed modes), with the complexity of the spectrum instead coming from large rotational splittings from its fast rotation rate.

We suggest instead the possibility that Zvrk is a post-dredge-up He WD--MS merger remnant which is on the CHeB phase, \textit{not} the RGB.
This addresses the aforementioned observations in the following ways:
\begin{enumerate}
    \item If Zvrk were a CHeB star, it would naturally have a large mixed-mode coupling and, thus, a CHeB-like spectrum (see also Section \ref{chebcoupling}).

    \item Our models of post-dredge-up CHeB remnants (Section \ref{overluminous}), like Zvrk, have radii approximately twice as large as those of typical CHeB stars.

    \item As demonstrated in Section \ref{chebabund}, post-dredge-up remnants are expected to be carbon-rich, similar to Zvrk.
    While the core dredge-up event at the helium flash is unlikely to match the measured value of $\mathrm{A}(\mathrm{Li})$, lithium enrichment may occur at merger \edit{\citep[as suggested by observations of luminous red novae]{kaminski2023lithium}.}
    \edit{Although some of our merger remnant models become very lithium-rich soon after merger (as we describe in Section \ref{chebabund}), future work should address whether this enhancement persists under more careful modeling.}

    \item As discussed in Section \ref{rotationsect}, merger remnants are \edit{likely} rapidly rotating.
\end{enumerate}

It is unclear at present whether He WD--MS mergers can reproduce these effects in the correct combination to match observations.
The mass $M\approx1.2M_\odot$ inferred by scaling relations is larger than those of our fiducial models ($M=0.8M_\odot$), possibly requiring even stronger core dredge-ups which probably occur for larger values of $M_{\mathrm{WD}}$ than we can model.
We point out that an extreme dredge-up event of this type probably modifies $\nu_{\mathrm{max}}$ \citep[in a still-contested manner;][]{viani2017changing,zhou2024does} and, thus, the accuracy of scaling relation-based values of $M$ and $R$.
The similarly behaving remnants of RG--He WD mergers \edit{\citep[e.g.,][]{zhang2013white}} may also possibly reproduce the properties of Zvrk.

\newNZR{Further complicating the picture, \citet{ong2024gasing} point out that a na\"ive identification of the double-ridged feature in Zvrk's \'echelle diagram with the usual $\ell=0$ and $2$ degrees implies an unusually small p-mode offset $\epsilon_p\sim0.25$.
This is too low to be consistent with a low-mass RG of any canonical evolutionary state: none of the observed stars in Figure 10 of \citet{kallinger2012evolutionary} have $\epsilon_p<0.4$ (though CHeB stars \textit{do} have lower values of $\epsilon_p$ than do first-ascent RGs with comparable $\Delta\nu$).
Of course, if Zvrk is a post-dredge-up remnant, it may well be possible that it attains an unusual value of $\epsilon_p$.
While not theoretically characterized in this work, $\epsilon_p$ may turn out to be another observational diagnostic for He WD--MS remnants.}


While it is beyond the scope of the present work, we encourage a more detailed investigation to determine whether this hypothesis can explain Zvrk's large radius, oscillation spectrum, surface abundances, and rapid rotation in a quantitative and self-consistent way.

\subsection{Other potential post-dredge-up merger remnants}

\newNZR{On the $\Delta\nu$--$\Delta\Pi$ diagram, \citet{mosser2014mixed} identify several RGs which lie near, but slightly leftward, of the red clump (see their Figure 1).
While \citet{mosser2014mixed} argue that these stars have recently undergone helium subflashes, we suggest that they might be post-dredge-up remnants.
As we show in Section \ref{chebdeltapg}, post-dredge-up remnants are also expected to have values of $\Delta\Pi$ comparable to those of normal CHeB stars, but smaller values of $\Delta\nu$ on account of their larger radii.}

\newNZR{Recently, a small number of highly carbon-deficient \edit{red} giants has been discovered and characterized as a distinct class \edit{with a few common properties} \citep{bidelman1973brighter,bond2019carbon,maben2023asteroseismology}.
These objects are almost all in the CHeB phase \edit{(as implied by their g-mode period spacings)}, and many are also lithium-rich and overluminous compared to the usual red clump \citep{maben2023asteroseismology}.
However, the He WD--MS merger scenario predicts a carbon \textit{enrichment} and thus fails to explain the abnormally \textit{low} carbon abundances in these stars.}

\section{Discussion and future prospects} \label{discussion}




\subsection{Progenitors and rates} \label{rates}

\begin{figure}
    \centering
    \includegraphics[width=0.48\textwidth]{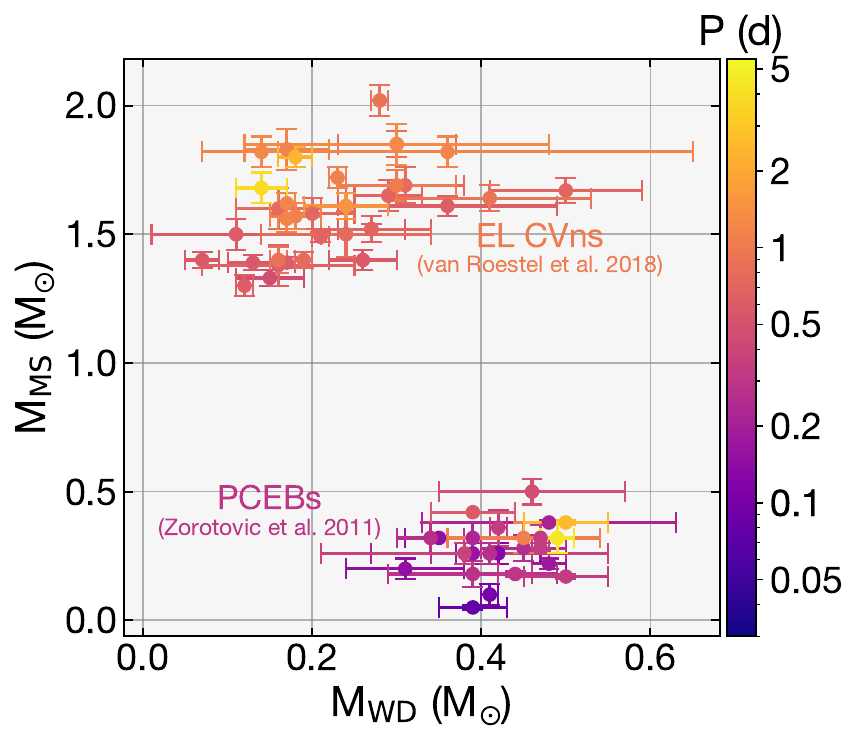}
    \caption{Parameters for a collection of known pre-CVs which may contain He WDs ($M_{\mathrm{WD}}\leq0.5M_\odot$) taken from \citet{zorotovic2011post} for PCEBs and \citet{van2018discovery} for EL CVns. Color indicates the orbital periods of each system.}
    \label{fig:precvs}
\end{figure}

While \edit{very few} CVs with He WD accretors have been discovered (to our knowledge), CV progenitors (known as pre-CVs) \textit{have} been, some shown in Figure \ref{fig:precvs}.
As their name suggests, such systems are expected to eventually initiate (possibly unstable) mass transfer after a combination of magnetic braking and gravitational radiation tighten their orbits sufficiently.

\citet{zorotovic2011post} compile a catalog of post-common envelope binaries (PCEBs), WD--MS binary systems, some of which may contain He WD components, identified in this work as those with $M_{\mathrm{WD}}<0.5M_\odot$ (see Figure \ref{fig:precvs}).
Most of the MS components are of relatively low mass, with $M_{\mathrm{MS}}\simeq0.3M_\odot$.
However, many of these systems contain fairly massive He WDs and could, upon merging, display fairly extreme versions of the asteroseismic and photometric merger remnant signatures we propose.

In later years, \citet{maxted2014cvn} identified a separate class of close detached binaries involving AF-type MS stars orbiting low-mass ($\simeq0.2M_\odot$) proto-He WDs, with \citet{van2018discovery} later measuring $M_{\mathrm{WD}}$ and $M_{\mathrm{MS}}$ for $36$ such systems (see Figure \ref{fig:precvs}).
Due to the low masses of the proto-WD components, EL CVns are likely formed via stable mass transfer rather than common-envelope events \citep[which would likely result in merger;][]{chen2017formation}.
The EL CVns in Figure \ref{fig:precvs} will likely merge when the stellar component is either a MS star or a subgiant \citep{lagos2020most}.
\newNZR{In the He WD--subgiant case, the He WD is expected to merge with the subgiant core and produce a similar, low-entropy-core remnant as in the He WD--MS case explored in this work.
However, He WD--subgiant mergers may also have their own distinctive signatures, and a detailed investigation of the associated remnants is probably warranted.}


Based on MESA models of merger remnants \citep{zhang2017evolution}, \citet{zhang2018evolution} perform a population synthesis calculation to determine how much these sorts of merger remnants contribute to the population of single He WDs, which are not a natural outcome of isolated stellar evolution \edit{\citep[see also][]{zorotovic2017origin}}.
Assuming a star-formation rate of $5M_\odot\,\mathrm{yr}^{-1}$, their calculation implies a Galactic formation rate of merger remnants $\sim0.02\,\mathrm{yr}^{-1}$ (about half of which fizzle out into He WDs before starting helium burning), with a factor-of-a-few uncertainty when the Reimers wind parameter is varied \citep{reimers1975circumstellar,reimers1977absolute}.
If merger remnants continue to appear as RGs for $100T_{\mathrm{m}8}\,\mathrm{Myr}$, the Milky Way should contain $\sim2T_{\mathrm{m}8}\times10^6$ remnants at a given time.
Put another way, for a Kroupa initial mass function \citep{kroupa2001variation}, single low-mass RGs in the range $0.8M_\odot$ to $2.0M_\odot$ form at a rate $\sim0.3M_\odot\,\mathrm{yr}^{-1}$.
If isolated RGs have lifetimes $100T_{\mathrm{s}8}\,\mathrm{Myr}$, He WD--MS merger remnants should make up roughly $\sim7\%\times(T_{\mathrm{m}8}/T_{\mathrm{s}8})$ of all low-mass RGs.

\edit{The present work has focused on the case where the compact component is a He WD.
However, in principle, lower-mass CO WDs may also participate in mergers with MS stars for the same reasons.
In such cases, the MS component is expected to disrupt around and accrete onto the carbon--oxygen core and initiate shell burning, and the merger remnant is therefore likely to resemble an asymptotic giant branch star with an unusually cold core.
We defer a detailed investigation of this interesting possibility to a future work.}


\subsection{Merger transients and mass retention}

The observable signatures of He WD--MS merger remnants described in this paper are applicable long (at least a thermal time) after the merger.
However, the merger itself as well as any preceding novae events should produce observable transients and other emission, whose rates should be consistent with the population of merger remnants.



\citet{metzger2021transients} show that the merger itself likely produces a dusty transient similar to a luminous red nova \citep[e.g.,][]{kulkarni2007unusually} powered by recombination of the ejected material, and roughly estimate a Galactic rate $\sim0.1\,\mathrm{yr}^{-1}$.
This is larger than the rate predicted by \citet{zhang2018evolution} by a factor $\sim5$, although both estimates are subject to significant uncertainties.
From this rate, they estimate that the Milky Way contains $10^3$--$10^4$ remnants at the present day, which is much smaller than the $\sim10^6$ we estimate in Section \ref{rates}.
The reason is that their models predict the merger remnant only retains a small fraction $\sim \! 10\%$ of the hydrogen supplied by an already low-mass MS star, rather than all of it. This implies a much shorter post-merger lifetime $\sim10$--$100$ kyr.

However, \cite{metzger2021transients} investigates cases where the MS star has a low mass, $M_{\rm MS} \lesssim 0.5 M_{\rm WD}$.
As they point out, higher-mass MS stars will likely produce gravitationally unstable disks that accrete much more efficiently onto the WD, closer to our assumption of conservative mass accretion.
Ignition of hydrogen burning on the WD during the merger (not included in their simulations) may also cause the remnant to swell up and gain more mass from the disk.
An independent constraint on the population of He WD--MS merger remnants in the Milky Way may help distinguish these scenarios.

\subsection{Rapid rotation} \label{rotationsect}

\edit{While most RG envelopes rotate very slowly, spectroscopic \citep{carlberg2011frequency} and photometric \citep{ceillier2017surface} studies have indicated that $\sim2\%$ of RGs are rapidly rotating ($v\sin i\gtrsim10\,\mathrm{km}\,\mathrm{s}^{-1}$, or $v/v_{\mathrm{crit}}\gtrsim7\%$ for typical values $M\sim M_\odot$ and $R\sim10R_\odot$), likely as a result of stellar interactions such as mergers.
The He WD--MS merger remnants discussed in this work may also rotate rapidly enough for rotation measurements to serve as an orthogonal diagnosis for their binary origin.}

The merger occurs when the MS component of the close binary overflows its Roche lobe. Very roughly, this occurs at a semimajor axis $a \! \sim \! 3 \, R_{\mathrm{MS}} \sim 3 R_\odot(M_{\mathrm{MS}}/M_\odot)^{0.8}\mathrm{,}$ 
according to the traditional mass--radius scaling formula for the MS.
At the time of merger, the total binary has an orbital angular momentum
\begin{equation}
    L = M_{\mathrm{MS}}M_{\mathrm{WD}}\sqrt{\frac{Ga}{M}} 
\end{equation}
\noindent where $M_{\mathrm{tot}}=M_{\mathrm{MS}}+M_{\mathrm{WD}}$ (and we have assumed a circular orbit).
Then, assuming that no mass is lost during the merger, the final envelope spin rate of the resulting RG remnant is
\begin{equation} \label{rotationest}
    \begin{split}
        \Omega_{\mathrm{env}} &\sim 3.8\,\mu\mathrm{Hz}\times\left(\frac{\kappa}{0.2}\right)^{-1}\left(\frac{M_{\mathrm{MS}}}{0.5M_\odot}\right)^{0.4}\left(\frac{M_{\mathrm{WD}}}{0.3M_\odot}\right)\\
        \times&\left(\frac{M}{0.8M_\odot}\right)^{-1/2}\left(\frac{R}{20R_\odot}\right)^{-2}
    \end{split}
\end{equation}

\noindent where we have scaled the moment of inertia of the merger remnant's envelope to $\kappa=I/M_{\mathrm{MS}}R^2=0.2$ \citep[as in, e.g.,][]{bear2010spinning}.

This corresponds to rotation periods $\sim \! 20\,\mathrm{days}$, more than half the surface breakup frequency $\Omega_K=\sqrt{GM/R^3}\sim6.2\,\mu\mathrm{Hz}$, depending on the inflated radius of the remnant.
This is much faster than the rotation rate of typical RG stars at similar radii: rapid rotation \newNZR{(without associated radial velocity variability)} and associated magnetic activity can therefore help distinguish merger remnants.

However, Equation \ref{rotationest} gives an estimate of the envelope rotation rate very soon (roughly a thermal time) after the merger.
The subsequent evolution of the rotation \newNZR{profile} is strongly dependent on the physics of magnetic braking (which saps angular momentum from the system) and angular momentum transport (which couples the core and envelope rotation rates).
Both of these pieces of physics are not particularly well understood (especially in the context of fast-rotating RGs),
and models which incorporate both of these effects are necessary for predicting the long-term evolution of core and envelope rotation rates of merger remnants (Qian et al., in preparation).

\edit{Throughout this work, we have neglected rotation-induced mixing processes \citep{zahn1994rotation,talon1997rotational,mathis2004transport,zahn2010rapid,park2020horizontal} which may be important in a rapidly rotating merger remnant.
Our predictions for the amount of helium and other species mixed into the envelope and brought to the surface during a core dredge-up event should thus be considered lower limits.}


\subsection{Non-asymptotic effects on pulsations}

Both Equations \ref{dnu} and \ref{dpgformula} for $\Delta\nu$
and $\Delta\Pi$
rely on the \textit{asymptotic} approximation for stellar oscillations, i.e., that the radial wavelength of the oscillation is much smaller than the scale height of any structural variable.
While the asymptotic approximation is well-justified for most RGs \citep[see, e.g., the introductory discussion of][]{ong2020semianalytic}, merger remnants may possess sharp features in their profiles (glitches) which may cause departures from the asymptotic formulae.
Indeed, such glitches are observed in a small fraction of CHeB red giants \citep{vrard2021glitches}.

Although we eschew a comprehensive analysis of non-asymptotic effects in this work, we point out three possible glitches which may occur in He WD--MS merger remnants:
\begin{enumerate}
    \item Soon after merger, the Brunt--V\"ais\"al\"a frequency rises sharply with radius at the interface between the He WD and hydrogen-burning shell.
    In intermediate-mass MS stars, a similar spike in $N$ near the convective core can produce variations in $\Delta\Pi$ versus $P$ whose ``period'' in this space is a function of the buoyancy coordinate of the glitch \citep{miglio2008probing,pedersen2018shape}.
    Given the low-entropy state of the core and the short-lived nature of the entropy discontinuity, it remains to be seen whether this buoyancy feature is detectable.

    \item In merger remnants, more intense helium flashes quickly burn larger fractions of helium into carbon.
    This enhances the composition gradient between helium flash-processed material and outer layers of the helium core.
    This may produce an abnormally strong compositional peak in $N$ during the CHeB phase, which may manifest as an observable buoyancy glitch.
    
    \item At the He \textsc{i} and He \textsc{ii} ionization zones, the first adiabatic exponent dips abruptly, producing sharp features in the sound speed \citep[e.g.,][]{miglio2010evidence}.
    Notably, the amplitudes of these acoustic glitches increase with higher helium mass fraction $Y$ \citep{houdek2007asteroseismic}.
    During the CHeB phase of our models, $Y$ can be enhanced to extreme degrees (Section \ref{chebsupersect}), and it is possible that the effect of these acoustic glitches may be very strong.
\end{enumerate}

Detailed mode calculations are likely required to determine whether these glitches are observable, what their characteristics are in the oscillation spectrum, and to what extent they can be used to identify and characterize merger remnants.

\newNZR{Our predictions for the large frequency spacing $\Delta\nu$ relies on the scaling relations in Equation \ref{dnu}, which is known to require corrections when the outer layers deviate from homology to a calibration standard \citep{belkacem2013seismic,ong2019explaining}.
This may be slightly modified in merger remnants by changes in surface composition or rapid rotation, and should be investigated in the future.}

\subsection{Broader progress in binary interaction asteroseismology} \label{widerworld}

Due to the diversity of stellar interactions expected to occur in the field, other post-merger stellar structures and their asteroseismic signatures deserve future investigation, in particular:
\begin{itemize}
    \item \textit{RG--He WD mergers} likely produce unusual CHeB giants, which have been previously explored as a possible channel for producing certain classes of carbon stars \citep{izzard2007origin,zhang2013white,zhang2020population} \edit{and} as possible progenitor\edit{s} of the 8 UMi planetary system \citep{hon2023close} \edit{and CK Vulpeculae, a historical transient observed in the year 1670 which is now a bipolar nebula \citep{tylenda2024nova}}.
    The models of \citet{zhang2013white} suggest that these merger remnants behave like overluminous CHeB stars, similar to those described in Section \ref{overluminous} for the He WD--MS scenario (compare their Figure 4 to our Figure \ref{fig:hrd_flashclump}).

    \item \textit{CO WD--MS mergers} are likely to result from consequential angular momentum loss, particularly for lower-mass CO WDs.
    While existing \edit{modeling} literature typically focuses on progenitor systems' nova eruptions \citep{iben1996evolution,shara2010extended,kato2017recurrent} or their role in producing Type Ia supernovae \citep{kovetz1994accretion,cassisi1998hydrogen,newsham2013evolution,hillman2016growing,wang2018mass}, mergers of such remnants should become unusual asymptotic giant branch stars with highly degenerate cores \citep{cassisi1998hydrogen,piersanti2000hydrogen,wolf2013hydrogen}, with possibly observable consequences.

    \item \textit{CO WD--He WD mergers} are the likely progenitors of \textit{R Coronae Borealis stars} (\textit{R Cor Bor} stars), which are yellow supergiants consisting of a carbon--oxygen core surrounded by a helium envelope inflated by shell burning \edit{\citep{clayton2007very,menon2013reproducing}}.
    R Cor Bor stars are known to pulsate at periods between $40$ and $100\,\mathrm{d}$ \citep{lawson1996observational,karambelkar2021census}, making asteroseismology a promising tool for probing their internal structures \citep{wong2023asteroseismological}.
    
    \item Last year, \citet{bellinger2023potential} identified asteroseismology as a tool for testing the post-MS merger channel for producing \textit{blue supergiants}, finding
    that the g-mode period spacing $\Delta\Pi$ constrains their formation channels.
    In a parallel observational study using TESS, \citet{ma2023variability} discovered \edit{a} peculiar but universal low-frequency ($f\lesssim2\,\mathrm{d}^{-1}$) photometric power excess, although the physical \edit{nature of these oscillations} remains unclear\edit{, and the authors were unable to observe individual modes or measure $\Delta\Pi$}.
\end{itemize}

Binary interaction asteroseismology is a technique at its infancy, with likely many more fruitful directions.

\section{Summary} \label{conclusion}

In this work, we presented detailed models of merger remnants of He WD--MS mergers.
Merger remnants quickly initiate hydrogen shell burning and become unusual giant stars which may hide inside the RG population.
However, they exhibit a number of unique signatures which may be used to distinguish them.
In summary, during hydrogen shell-burning (RGB), merger remnants:

\begin{enumerate}
    \item are over-inflated at a given core mass.

    \item depart from the standard degenerate sequence on the asteroseismic $\Delta\nu$--$\Delta\Pi$ diagram.
    Asteroseismology can thus identify remnants whose mixed-mode coupling is sufficiently strong.

    \item undergo delayed helium flashes, and attain higher luminosities at the tRGB than do single RGs.
    
\end{enumerate}

During helium core burning (CHeB), remnants of mergers involving lower-mass He WDs:
\begin{enumerate}
    \item attain core masses which are essentially typical for single CHeB stars.
    
    \item are strong candidates for the undermassive red clump stars discovered by \citet{li2022discovery} and \citet{matteuzzi2023red}.
\end{enumerate}

Remnants of mergers involving higher-mass He WDs dredge up a significant fraction (up to $\sim0.1M_\odot$) of helium into the envelope.
During core helium burning, these post-dredge-up merger remnants:

\begin{enumerate}
    \item have significantly larger radii and luminosities than single stars on the red clump.

    \item exhibit a wider range of asteroseismic g-mode period spacings $\Delta\Pi$ than do typical stars on the red clump.

    \item attain abnormally strong degrees of asteroseismic mixed-mode coupling towards the end of CHeB.

    \item are enriched in ${}^{12}\mathrm{C}$, as well as ${}^{18}\mathrm{O}$ and ${}^{22}\mathrm{Ne}$.

    \item may already have been discovered.
    The rapidly rotating RG discovered by \citet{ong2024gasing} (``Zvrk'') has many of the predicted properties of this type of merger remnant.
\end{enumerate}

Observational probes of these merger remnants can constrain the He WD--MS merger process at the population level.
In turn, this may provide additional confirmation of the consequential angular momentum loss hypothesis and white dwarf mass problem for CVs.\\

\noindent We thank Kareem El-Badry, Joel Ong, Marc Hon, and Yaguang Li for their insightful comments and thorough reading of the manuscript, as well as Lars Bildsten for useful discussions. We also acknowledge Masao Takata, Andrew Casey, Tim Bedding\edit{, and Ken Shen} for their helpful remarks\edit{, as well as the anonymous referee whose report increased the clarity of the final manuscript}.
N.Z.R. acknowledges support from the National Science Foundation Graduate Research Fellowship under Grant No. DGE‐1745301.
\edit{All stellar evolution calculations were performed on the Wheeler cluster at Caltech, which was supported by the Sherman Fairchild Foundation and by
Caltech.}\\


{\large\textit{Software:}} Modules for Experiments in Stellar Astrophysics \citep{paxton2010modules,paxton2013modules,paxton2015modules,paxton2018modules}

\bibliography{sample631}{}
\bibliographystyle{aasjournal}

\appendix

\section{\edit{A.} Dependence on the cooling age of the white dwarf} \label{coolingage}

Pre-merger, all of the merger models discussed in the main text use a He WD which has been cooled until it achieves a luminosity $L_{\mathrm{WD}}=10^{-4.0}L_\odot$.
This relatively low luminosity is chosen to explore the limiting case of a very degenerate helium core.
In our models, this luminosity corresponds to relatively long He WD cooling ages $5 \, {\rm Gyr} \lesssim t_{\mathrm{cool}}\lesssim7\,\mathrm{Gyr}$, with a fixed $L_{\mathrm{WD}}$ corresponding to longer $t_{\mathrm{cool}}$ for He WDs with higher masses $M_{\mathrm{WD}}$ or more substantial atmospheres.
For comparison, note that the merger remnant models of \citet{zhang2017evolution} use $L_{\mathrm{WD}}=10^{-2.0}L_\odot$.

\begin{figure*}
    \centering
    \includegraphics[width=\textwidth]{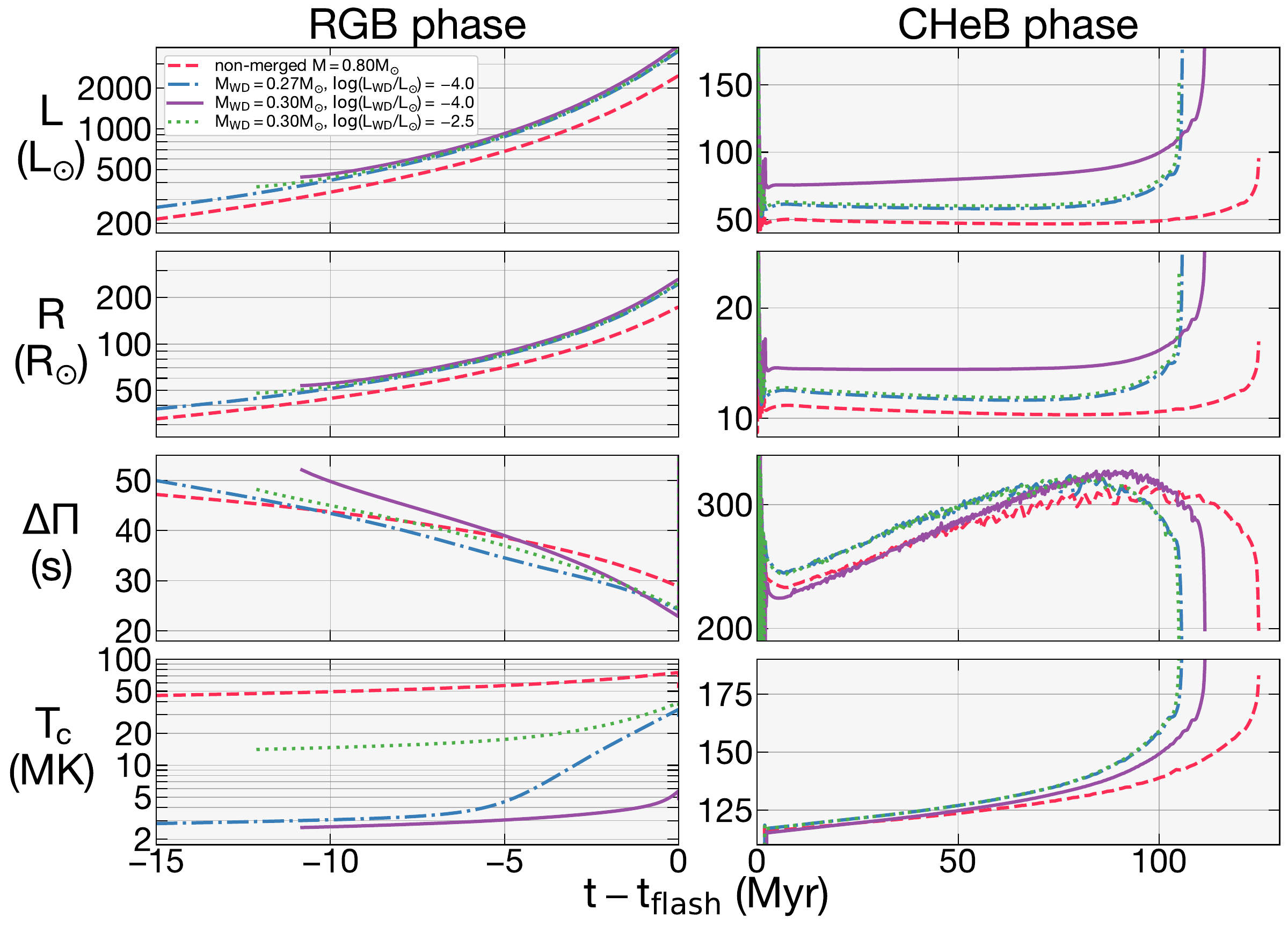}
    \caption{The total luminosity $L$, radius $R$, asteroseismic g-mode period spacing $\Delta\Pi$, and central temperature $T_c$ for selected models during the RGB (\textit{left}) and CHeB (\textit{right}) phases.
    A model with $M_{\mathrm{WD}}=0.30M_\odot$ but a brighter He WD ($\log(L_{\mathrm{WD}}/L_\odot)=-2.5$) is included.}
    \label{fig:logLinit_over_time_test}
\end{figure*}

In this Appendix, we discuss the effect of varying $\log(L_{\mathrm{WD}}/L_\odot)$.
Figure \ref{fig:logLinit_over_time_test} shows the evolution of some selected selected quantities for four models (the first three of which also appear in the main text):
\begin{enumerate}
    \item A non-merged model, which has a ``normal'' helium core on the RGB which is close to isothermal with the hydrogen-burning shell.

    \item A merger remnant with $\log(L_{\mathrm{WD}}/L_\odot)=-4.0$ and $M_{\mathrm{WD}}=0.30M_\odot$.

    \item A merger remnant with $\log(L_{\mathrm{WD}}/L_\odot)=-4.0$ and a \textit{lower\edit{-}mass} $M_{\mathrm{WD}}=0.27M_\odot$.

    \item A merger remnant with a \textit{higher pre-merger He WD luminosity} $\log(L_{\mathrm{WD}}/L_\odot)=-2.5$ and $M_{\mathrm{WD}}=0.30M_\odot$.
    This value of $\log(L_{\mathrm{WD}}/L_\odot)$ corresponds to a He WD cooling age $t_{\mathrm{cool}}\approx550\,\mathrm{Myr}$.
\end{enumerate}
All models have total masses of $M=0.80M_\odot$ and WD atmospheres of mass $10^{-4}M_\odot$, and respectively.

As described in Section \ref{rgb}, merger remnants during the RGB will start with initially low-entropy cores, but their entropies will gradually grow due to heat diffusion and deposition of higher-entropy helium resulting from hydrogen burning.
The degree of entropy deficit in the core therefore results from a combination of the entropy of the original He WD as well as its mass, which determines its total heat capacity).
Because increasing $\log(L_{\mathrm{WD}}/L_\odot)$ and decreasing $M_{\mathrm{WD}}$ affect this core entropy deficit in the same way, merger remnants should be affected by increases in $\log(L_{\mathrm{WD}}/L_\odot)$ and decreases in $M_{\mathrm{WD}}$ in similar ways.

Figure \ref{fig:logLinit_over_time_test} shows the evolution of the models listed above. They all have comparable radii on the RGB, but $\Delta\Pi$ (reflecting the internal thermal structure of the core) varies somewhat between them.
As can be seen on the \textit{bottom left} panel of Figure \ref{fig:logLinit_over_time_test}, the core temperature $T_c$ of the $M_{\mathrm{WD}}=0.27M_\odot$, $\log(L_{\mathrm{WD}}/L_\odot)=-4.0$ model initially behaves very similarly to the $M_{\mathrm{WD}}=0.30M_\odot$, $\log(L_{\mathrm{WD}}/L_\odot)=-4.0$ model, since not enough time has yet elapsed for conduction to significantly modify its temperature.
Later on, $T_c$ in the $M_{\mathrm{WD}}=0.27M_\odot$ model takes a sharp upturn to more closely resemble the warmer $M_{\mathrm{WD}}=0.30M_\odot$, $\log(L_{\mathrm{WD}}/L_\odot)=-2.5$ model.
By the helium flash, these two models have very similar $\Delta\Pi$ and temperature profiles.

Once the helium flash occurs, variations in the temperature profile due to a finite thermal conductivity in the core are erased entirely.
During the CHeB phase, the $M_{\mathrm{WD}}=0.27M_\odot$, $\log(L_{\mathrm{WD}}/L_\odot)=-4.0$ and $M_{\mathrm{WD}}=0.30M_\odot$, $\log(L_{\mathrm{WD}}/L_\odot)=-2.5$ models are essentially identical (\textit{right} panels of Figure \ref{fig:logLinit_over_time_test}).
Both models evolve significantly differently than the $M_{\mathrm{WD}}=0.3M_\odot$, $\log(L_{\mathrm{WD}}/L_\odot)=-4.0$ model (which had a larger entropy deficit on the RGB).

In summary, decreasing the cooling age of the merging He WD has a very similar effect to slightly decreasing its mass. Hence, the merger models in the main text are expected to behave similarly to merger models with younger and slightly more massive WDs.

\section{\edit{B.} Thermal timescale of the core} \label{coreth}

In this Appendix, we estimate the timescale $\tau_{\mathrm{th,core}}$ for a cold, highly degenerate remnant core to thermalize with a hot burning shell.
In this environment, electron-mediated heat conduction is the dominant heat transport mechanism.

The resulting heat flux $\Vec{q}$ takes the form
\begin{equation} \label{heat}
    \Vec{q} = -\alpha_{\mathrm{cond}}\rho T\nabla T
\end{equation}

\noindent where
\begin{equation}
    \alpha_{\mathrm{cond}} \approx 2.44\times10^3AZ^{-2}\,\mathrm{cm}^4\,\mathrm{s}^{-3}\,\mathrm{K}^{-2}\mathrm{,}
\end{equation}

\noindent see \citet{mestel1950thermal}.
The heat capacity per unit volume is dominated by the non-degenerate ions:
\begin{equation} \label{cv}
    c_v \approx \frac{\varepsilon}{T} \approx \frac{3\rho k_B}{2Am_p}
\end{equation}

In the absence of heat sources, the continuity equation for energy density $\varepsilon$ within the interior of the WD is thus given by
\begin{equation} \label{cont}
    \frac{\partial\varepsilon}{\partial t} = -\nabla\cdot\Vec{q}
\end{equation}

Substituting Equations \ref{heat} and \ref{cv} into Equation \ref{cont} yields the following nonlinear heat diffusion equation:
\begin{equation} \label{nonlinear}
    \frac{\partial T}{\partial t} = \frac{\kappa}{r^2}\frac{\partial}{\partial r}\left(r^2T\frac{\partial T}{\partial r}\right)\mathrm{,}
\end{equation}

\noindent cf. \citet{mestel1952theory}.
In writing Equation \ref{nonlinear}, we have assumed spherical symmetry, and that $\rho$ does not vary with temperature.
The diffusion coefficient $\kappa$ is given by
\begin{equation}
    \kappa = \frac{2Am_p\alpha_{\mathrm{cond}}}{3k_B}
\end{equation}

At early times, the core is essentially isothermal at a low temperature $T_0\ll T_{\mathrm{shell}}\simeq3\times10^7\,\mathrm{K}$.
Thermal contact with the hot hydrogen-burning shell at the outer boundary sets up a temperature gradient $\partial T/\partial r\sim T_{\mathrm{shell}}/R_c$.
The natural length scale of the problem is $R_c$.
By Equation \ref{nonlinear}, the core therefore thermalizes on a timescale $\tau_{\mathrm{th,core}}$ given roughly by
\begin{equation}
    \frac{1}{\tau_{\mathrm{th,core}}} \sim \frac{\kappa T_{\mathrm{shell}}}{R_c^2}
\end{equation}

\noindent or
\begin{equation} \label{timescale}
    \tau_{\mathrm{th,core}} \sim \frac{3k_B}{2Am_p\alpha_{\mathrm{cond}}}\frac{R_c^2}{T_{\mathrm{shell}}} \approx 80\,\mathrm{Myr}\times\left(\frac{R_c}{0.035R_\odot}\right)^2\left(\frac{T_{\mathrm{shell}}}{3\times10^7\,\mathrm{K}}\right)^{-1}\left(\frac{Z}{2}\right)^2\left(\frac{A}{4}\right)^{-2}
\end{equation}

A similar timescale to $\tau_{\mathrm{th,core}}$ has been previously derived by \citet{shen2009helium}.
The scaling relation in Equation \ref{timescale} is normalized to a typical He WD in thermal contact with a typical hydrogen-burning shell.
This timescale is comparable to the length of the remnant's RGB phase---the longest RGB phase leading to helium ignition in our merger remnant models lasts $\simeq180\,\mathrm{Myr}$ (for a He WD mass $M_{\mathrm{WD}}=0.20M_\odot$).
Heat conduction thus cannot completely destroy the low-entropy state of the core quickly enough to erase its long-term effects on the appearance and evolution of the remnant, especially for higher values of $M_{\mathrm{WD}}$.


\section{\edit{C.} Brunt--V\"ais\"al\"a frequency in degenerate helium cores} \label{degenbrunt}

In this Appendix, we derive the Brunt--V\"ais\"al\"a frequency $N$ in the part of the g-mode cavity which lies within the helium core of a star on the RGB.
This region is characterized by degenerate electrons which dominate the pressure support and non-degenerate ions which dominate the heat capacity.

Following \citet{brassard1991adiabatic}, in the absence of composition gradients, $N$ can be written as
\begin{equation}
    N^2 = N_0^2\frac{\chi_T}{\chi_\rho}\left(\nabla_{\mathrm{ad}} - \nabla\right)
\end{equation}

\noindent where
\begin{subequations}
    \begin{gather}
        \chi_T = \left(\frac{\partial\ln p}{\partial\ln T}\right)_\rho \\
        \chi_\rho = \left(\frac{\partial\ln p}{\partial\ln\rho}\right)_T \\
        \nabla_{\mathrm{ad}} = \left(\frac{\mathrm{d}\ln T}{\mathrm{d}\ln p}\right)_{\mathrm{ad}} \\
        \nabla = \frac{\mathrm{d}\ln T}{\mathrm{d}\ln p}
    \end{gather}
\end{subequations}

\noindent and we have defined
\begin{equation}
    N_0^2 = \frac{\rho g^2}{p}
\end{equation}

\noindent where roughly $N_0\!\simeq\!\sqrt{GM_c/R_c^3}$ is comparable to the dynamical frequency of the core.

In degeneracy-supported matter, $\chi_\rho\!\approx\!5/3$ and $\nabla_{\mathrm{ad}}\!\approx\!0.4$ are basically constant.
In WDs, the pressure has two contributions from the degenerate electrons (which dominate the total pressure) and non-degenerate ions (which carry the temperature dependence):
\begin{equation}
    p = \frac{2}{5}n_eE_F + n_ik_BT
\end{equation}

\noindent where $n_e=Z\rho/Am_p$ and $n_i=\rho/Am_p$ are the electron and ion number densities, respectively, and $E_F$ is the electronic Fermi energy. 
Roughly
\begin{subequations}
    \begin{gather}
        p \approx \frac{2}{5}\frac{Z\rho E_F}{Am_p} \\
        \left(\frac{\partial p}{\partial T}\right)_\rho \approx \frac{\rho k_B}{Am_p}
    \end{gather}
\end{subequations}

Then
\begin{equation}
    \chi_T = \frac{T}{p}\left(\frac{\partial p}{\partial T}\right)_\rho = \frac{5}{2Z}\frac{k_BT}{E_F}\mathrm{.}
\end{equation}

Therefore, for mostly degenerate matter,
\begin{equation}
    N^2 \approx N_0^2\frac{k_BT}{ZE_F}\left(1 - \frac{5}{2}\nabla\right)\mathrm{.}
\end{equation}

\section{\edit{D. Post-helium flash composition profiles}} \label{chebcomp}

\begin{figure*}
    \centering
    \includegraphics[width=\textwidth]{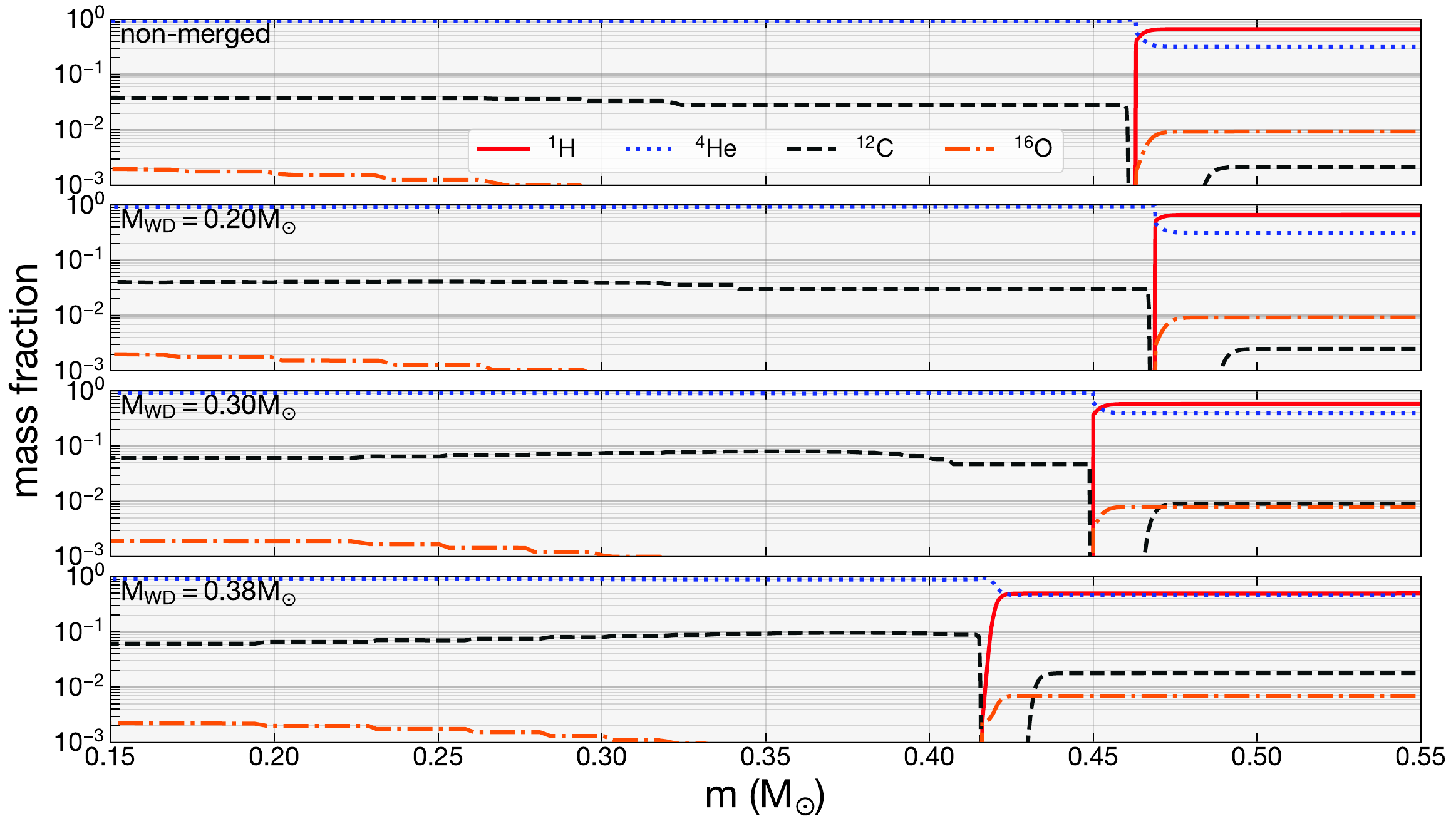}
    \caption{\edit{Composition profiles for the non-merged $M=0.80M_\odot$ and selected merger remnant models at the beginning of CHeB (a few megayear after the helium flash).}}
    \label{fig:interface_composition_1}
\end{figure*}

\edit{Figure \ref{fig:interface_composition_1} compares the composition profiles for various merger remnants at the beginning of CHeB, soon after the helium flash.}
\edit{As explained in Section \ref{chebsupersect}, merger remnants involving more massive He WDs (higher $M_{\mathrm{WD}}$) convert a larger fraction of their core helium into carbon during the helium flash.}
\edit{Moreover, our merger remnant models with $M_{\mathrm{WD}}\geq0.27M_\odot$ (e.g., the \textit{bottom} two panels of Figure \ref{fig:interface_composition_1}) undergo core dredge-up events which mix some of the helium core into the envelope.}
\edit{Such merger remnants therefore begin CHeB with lower core masses and possess envelopes which are dramatically enhanced in species such as helium and carbon (see Section \ref{chebabund}).}

\end{document}